\documentclass[pdflatex,sn-nature]{sn-jnl}

\usepackage{graphicx}%
\usepackage{multirow}%
\usepackage{amsmath,amssymb,amsfonts}%
\usepackage{amsthm}%
\usepackage{mathrsfs}%
\usepackage[title]{appendix}%
\usepackage{xcolor}%
\usepackage{textcomp}%
\usepackage{manyfoot}%
\usepackage{booktabs}%
\usepackage{algorithm}%
\usepackage{algorithmicx}%
\usepackage{algpseudocode}%
\usepackage{listings}%

\usepackage{mathtools} 
\usepackage{hyperref}

\theoremstyle{thmstyleone}%
\theoremstyle{thmstyletwo}%

\theoremstyle{thmstylethree}%
\begin{document}

\title[Article Title]{Multivariate linear regression without prior assumptions}


\author[1]{\fnm{Mayank S.K.} \sur{Gupta}}\email{mayankskg07@gmail.com}

\author[1]{\fnm{Deepanjhan} \sur{Das}}\email{deepanjhan.iitm22@gmail.com}

\author*[1]{\fnm{Arun K.} \sur{Tangirala}}\email{arunkt@iitm.ac.in}

\author*[1]{\fnm{Shankar} \sur{Narasimhan}}\email{naras@iitm.ac.in}

\affil[1]{\orgdiv{Department of Chemical Engineering}, \orgname{Indian Institute of Technology Madras}, \orgaddress{\street{Sardar Patel Road}, \city{Chennai}, \postcode{600036}, \state{Tamil Nadu}, \country{India}}}


%
%
\abstract{Recovering the linear relationships that govern a system from noisy measurements is a basic task across the physical and engineering sciences. Because every measured variable may carry an unknown amount of noise, classical regression must commit in advance to a set of structural assumptions: ordinary least squares requires a declared input-output partition with input variables being noise-free, total least squares assumes equal noise variance across all variables, and generalized total least squares additionally requires the noisy-variable partition and variances to be known beforehand. Kalman~\cite{Kalman:1982} showed that any procedure returning a unique linear model from inexact data must rest on such unverifiable a priori assumptions -- ``prejudices'' -- that cannot be checked against the data itself, and that removing them leaves the identification problem fundamentally indeterminate. Whether these prejudices can instead be resolved directly from the data has remain unresolved. Here we show that an iterative generalized-eigenvalue algorithm, QZ-IPCA, recovers the noisy-variable partition, noise variances, number of linear relations, and regression coefficients of a multivariate linear system simultaneously, using only the raw data. Across all possible exhaustive noise configurations of a five-variable benchmark network, QZ-IPCA correctly identifies model structure and recovers coefficients with error below 6.4\%. It outperforms ordinary least squares even when given the best partition, and succeeds in rank identification precisely where standard total least squares falls once noise variances differ across variables. These results show that the assumptions conventionally required for multivariate regression are not necessary, recasting model identification as a problem solvable from data geometry alone. We anticipate this assumption-free approach to extend naturally to other settings, such as sensor networks and industrial process monitoring, where the boundary between signal and noise is unknown in advance. More generally, it suggests that other classical estimation problems whose textbook solutions rest on similar prior assumptions may be reformulated to let the data determine its own structure.}


\maketitle

%
\section{Introduction} \label{sec:intro}
Every physical or engineering system that can be measured is, at some level, governed by relationships that hold exactly among its variables. For instances, such systems include mass and energy balances in a chemical plant~\cite{ShankarSir:1999}, conservation laws in a sensor network, calibration curves relating an instrument's reading to the quantity it measures~\cite{Brown:1982}. Recovering these relationships from data is one of the oldest problems in quantitative science, and it remains open in a precise sense, that is, every measured variable carries some amount of noise, and once more than one variable is imprecisely known, the covariance matrix computed from noisy data does not by itself reveal the relationships that produced it. The same matrix is compatible with many different combinations of which variables are corrupted, how large their noise are, and how many linearly independent relations hold among them.

For more than a century, resolving this ambiguity has required committing, in advance and independently of the data, to an assumption about which variables carry the noise. Ordinary least squares (OLS) declares one subset of variables noise-free ``input'' and forces all the noise onto the remainder~\cite{Kalman:1982,Los:1989,Hastie:2009}, called the ``output''. Total least squares (TLS), developed from Pearson's geometric formulation of principal component analysis (PCA)~\cite{Pearson:1901} and placed on firm numerical foundation via singular value decomposition (SVD)~\cite{Golub:1980,Huffel:1991}, in its classical formulation treats every variable as equally noisy. Generalized total least squares (GTLS) extends this further, but only if the noisy/non-noisy partition and the noise variances are supplied beforehand~\cite{Huffel:1989}. Kalman showed that this is not a shortcoming of any one method but a structural feature of the problem itself. Any procedure that returns a single, unique linear model from an inexact covariance matrix must rest on such an unverifiable a priori choice, which he termed a ``prejudice'' precisely because it cannot be checked against the data it is meant to explain~\cite{Kalman:1982}, a result later formalized for the classical estimators by Los~\cite{Los:1989}. The choice among OLS, TLS, and their variants is therefore itself a modeling decision made before any data is examined, and an incorrect choice biases the resulting model in ways the data alone cannot reveal.

\begin{figure*}[!htbp]
    \centering
    \includegraphics[width=\textwidth]{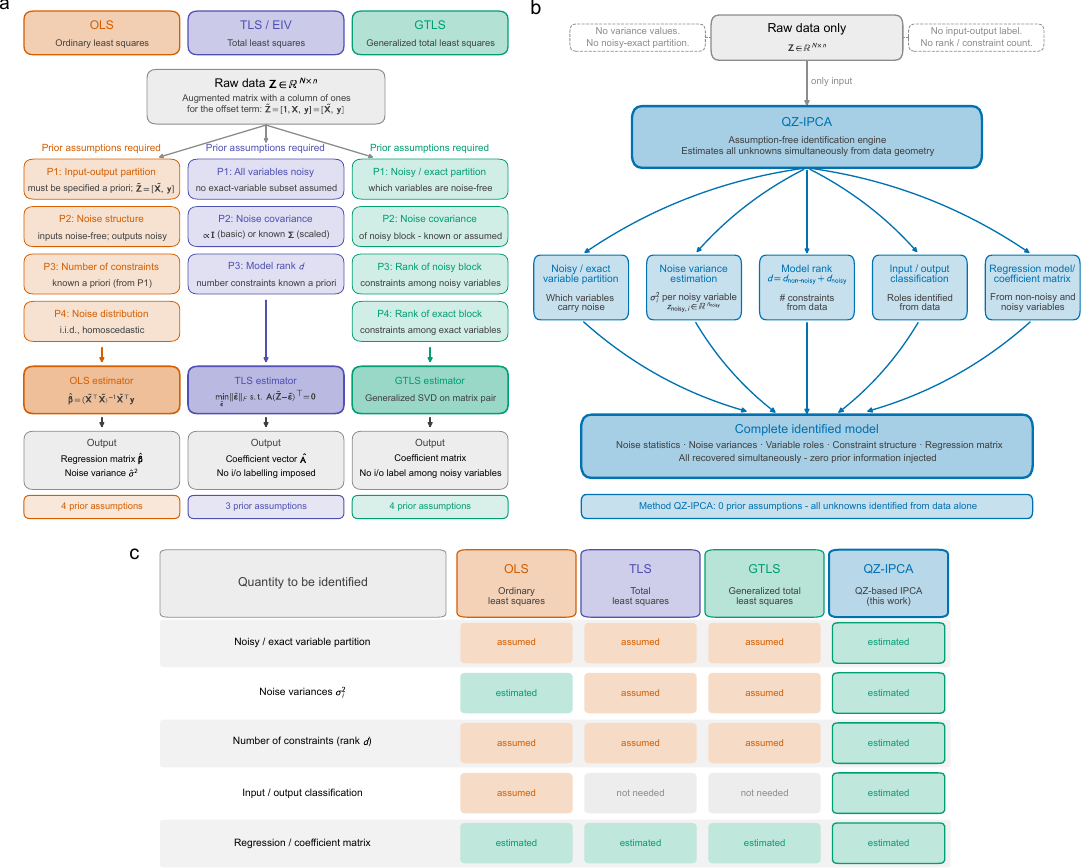}
    \caption{\textbf{Conventional identification methods require prior structural knowledge; QZ-IPCA requires none.} (\textbf{a}) Ordinary least squares (OLS)~\cite{Kalman:1982,Los:1989,Hastie:2009}, total least squares (TLS/EIV)~\cite{Pearson:1901,Golub:1980,Huffel:1991}, and generalized total least squares (GTLS)~\cite{Huffel:1989} each require a distinct set of prior structural assumptions, i.e., prejudices, before estimation can proceed. In all three cases the noise structure and number of linear constraints (model rank) must also be specified a priori. (\textbf{b}) QZ-IPCA accepts only the raw data matrix $\mathbf{Z}\in\mathbb{R}^{N\times n}$ with no variable labels, no noise model, and no rank specification, and recovers all unknowns simultaneously: the noisy/exact variable partition, per-variable noise variances, model rank, input/output roles, and the regression coefficient matrix. (\textbf{c}) Summary of which quantities are assumed versus estimated by each method. Across every conventional framework, at least three structural quantities must be assumed; QZ-IPCA estimates all of them from data geometry alone. Note that while OLS estimates the output noise variance $\sigma^2$, this is conditional on the input–output partition and the exact-input assumption having been pre-specified. It is a residual computation within an already-committed model frame, not a free identification of the noise structure.}
    \label{fig1:conventional_vs_qz}
\end{figure*}

Here we resolve this longstanding problem. We show that the noisy/non-noisy variable partition, the estimation of noise variances, the number of independent relations, and the relations themselves can be recovered simultaneously and directly from an unlabeled data matrix, with none of these quantities specified in advance. Our method, QZ-IPCA, combines principal component analysis (PCA) with an iterative, QZ-decomposition-based generalized eigenvalue solver to resolve the full identification problem from the data geometry alone. This subsumes OLS, TLS, and GTLS as special cases of a single underlying procedure rather than requiring a user to choose among them (cf. Figure~\ref{fig1:conventional_vs_qz}).

For demonstration purpose, we exhaustively applied the proposed algorithm across all 32 possible noise configurations of a five-variable benchmark network. QZ-IPCA correctly identifies the number and type of governing relations in every case and recovers regression coefficients with less than $6.4\%$ error, outperforming OLS even when OLS is given its best possible input-output partition. It further succeeds in rank identification precisely where standard TLS fails once noise variances differ across variables. This accuracy holds across a wide range of sample sizes and signal-to-noise ratios (SNRs), and is unaffected by whether the underlying process variables are excited persistently by Gaussian, sub-Gaussian, or super-Gaussian fluctuations. These results indicate that the assumptions long treated as unavoidable prejudices for multivariate linear regression are, in fact, unnecessary. The structure that Kalman~\cite{Kalman:1982} argued could only be imposed by prejudice can instead be read directly from the data.

\section{Results} \label{sec:results}
QZ-IPCA eliminates all prior assumptions that conventional regression methods require. OLS~\cite{Kalman:1982,Los:1989,Hastie:2009}, TLS~\cite{Pearson:1901,Golub:1980,Huffel:1991}, and GTLS~\cite{Huffel:1989} each demand a distinct set of structural prerequisites, such as variable partitioning into input/output and noisy/non-noisy classes, noise variance values, and model rank, i.e., the number of constraints. Before estimation can proceed, a minimum of three such quantities are needed to be assumed or specified a priori (Figure~\ref{fig1:conventional_vs_qz}\textbf{a},\textbf{c}). Our proposed unified identification framework, QZ-IPCA on the other hand, accepts only raw unlabeled data matrix $\mathbf{Z} \in \mathbb{R}^{N\times n}$ and recovers all of these quantities simultaneously: the noisy and noise-free variable classification, per-variable noise variance values, number of deterministic and stochastic constraints (the model rank), constraint matrix, input-output partition, and the regression coefficient matrix as also summarized in Figure~\ref{fig1:conventional_vs_qz}\textbf{b}. The sequential algorithmic pipeline that realizes this through PCA and iterative generalized eigenvalue decomposition is described in the Methods and summarized in Extended Data Figs. 1 and 2.

\subsection{QZ-IPCA correctly identifies model structure across all 32 noise topologies}
\begin{figure*}[!htbp]
    \centering
    \includegraphics[width=\textwidth]{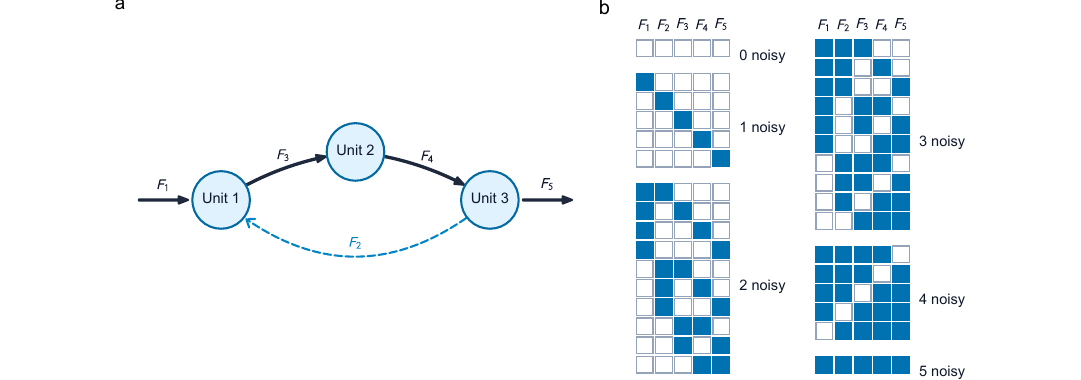}
    \caption{\textbf{Benchmark flow network and the exhaustive space of noise configurations.} (\textbf{a}) A five-variable flow network~\cite{ShankarSir:2008} comprising three process units and five flow variables ($F_1$--$F_5$), governed by three linear balance relations. This system serves as the validation benchmark for all 32 noise configurations studied. (\textbf{b}) Complete combinatorial space of noise assignments across the five variables, yielding $2^5 = 32$ distinct cases grouped by the number of noisy variables ($0$ through $5$). Left and right panels show cases with zero to two, and three to five noisy variables, respectively. Filled blue cells indicate noisy variables; empty cells indicate noise-free variables. The proposed method requires no prior specification of noise status and correctly identifies all $32$ configurations from data alone.}
    \label{fig2:flow_network_cases}
\end{figure*}

We evaluated QZ-IPCA on a five-variable flow network governed by three linear conservation relations as presented in Figure~\ref{fig2:flow_network_cases}\textbf{a}, which provides a structurally non-trivial but exactly characterized benchmark. The complete space of noise assignments across five binary-classified variables (either non-noisy or noisy) yields $2^5=32$ distinct configurations, spanning systems in which zero to all five flow variables carry measurement noise, as summarized in Figure~\ref{fig2:flow_network_cases}\textbf{b}. The set is exhaustive by construction. Every noise topology reachable within this network is included, and no information about the noise status or noise variance magnitude is supplied to the method in any case. 

\begin{figure*}[!htbp]
    \centering
    \includegraphics[width=\textwidth]{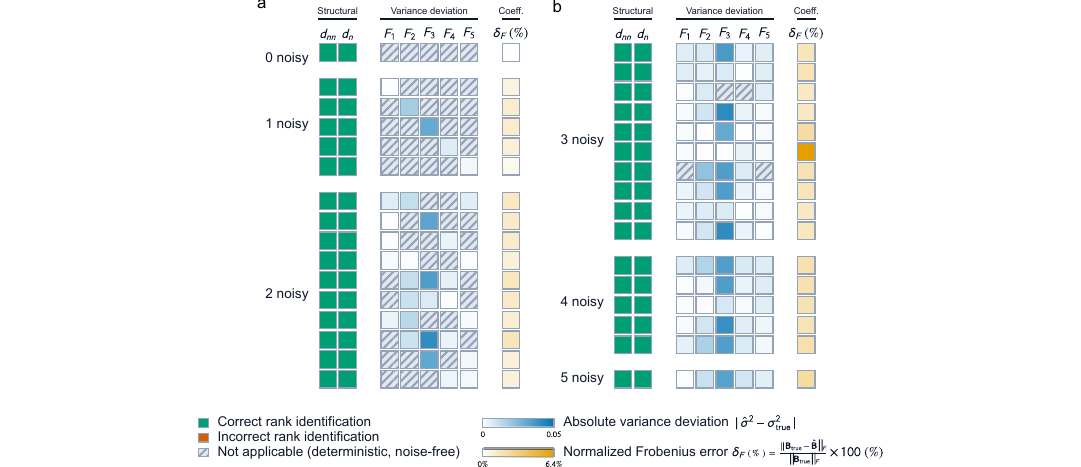}
    \caption{\textbf{QZ-IPCA correctly identifies rank, noise variances, and regression coefficients across all 32 simulation cases spanning 0 to 5 corrupted channels.} Performance metrics are systematically evaluated across simulation cases stratified by the number of corrupted channels: (\textbf{a}) low-density noise configurations (0--2 noisy variables) and (\textbf{b}) high-density noise configurations (3--5 noisy variables). The Structural columns report rank identification accuracy for `$d_{nn}$' (i.e., the number of non-noisy relations $d_{\text{non-noisy}}$) and `$d_n$' (i.e., the number of noisy relations $d_{\text{noisy}}$) with green denoting correct identification. The Variance deviation heatmaps display the absolute error in estimated noise variances $\left|\hat{\sigma}^2 - \sigma^2_{\text{true}}\right|$ across 5 flow variables ($F_1$--$F_5$); diagonally hatched cells indicate the identified non-noisy variables by the method prior to the iterative estimation stage, precluding further variance estimation. The `Coeff.' column reports the normalized Frobenius error $\delta_F(\%) = \| \mathbf{B}_{\text{true}} - \mathbf{\hat{B}} \|_F / \| \mathbf{B}_{\text{true}} \|_F \times 100$ of the estimated regression matrix. Rank is correctly identified in all cases; variance deviations remain below 0.05 throughout; and coefficient error stays within 6.4\% across all noise topologies.}
    \label{fig3:32_cases_results}
\end{figure*}

The success of QZ-IPCA in identifying the correct model rank, i.e., both the number of noise-free  (deterministic) relations $d_{\text{non-noisy}}$ and the number of noisy (stochastic) relations $d_{\text{noisy}}$, across all 32 configurations is presented in Figure~\ref{fig3:32_cases_results}. Absolute variance deviation $|\hat{\sigma}^2 - \sigma_{\text{true}}^2|$ remain below $0.05$ throughout, and the normalized Frobenius coefficient error $\delta_F$ stays within $6.4\%$ across every noise topology, including the structurally maximal Case 32 where measurements of all the five flow variables are noisy. Diagonally hatched cells in Figure~\ref{fig3:32_cases_results} indicate variables identified as non-noisy at the initial PCA stage. The variables are assigned zero noise variance deterministically and are excluded from iterative estimation, since their noise-free status, once established, precludes further stochastic characterization. Complete numerical values for all estimated quantities across all 32 cases, with 95\% bootstrap confidence intervals (CIs), are provided in Extended Data Tables 1--3. 

\subsection{QZ-IPCA outperforms OLS and standard TLS across representative noise regimes}
To benchmark coefficient recovery against established identification methods, we compared our proposed QZ-IPCA with OLS and standard TLS on three cases spanning the noise topology space (Figure~\ref{fig2:flow_network_cases}\textbf{b}): Case 1 ($d_{\text{non-noisy}}=3,$ $d_{\text{noisy}}=0$; all variables are noise-free), Case 19 ($d_{\text{non-noisy}}=1,$ $d_{\text{noisy}}=2$; $F_1,$ $F_2,$ $F_5$ are noisy), and Case 32 ($d_{\text{non-noisy}}=0,$ $d_{\text{noisy}}=3$; all variables are noisy). GTLS was excluded from this comparison because it requires the noise variances of all system variables along with the non-noisy and noisy variable partition as inputs~\cite{Huffel:1989}. In the absence of these quantities, which is precisely the operating condition under which an assumption-free comparison must be made, GTLS is operationally identical to standard TLS~\cite{Huffel:1991} and provides no additional capability. 

\begin{figure*}[!htbp]
    \centering
    \includegraphics[width=\textwidth]{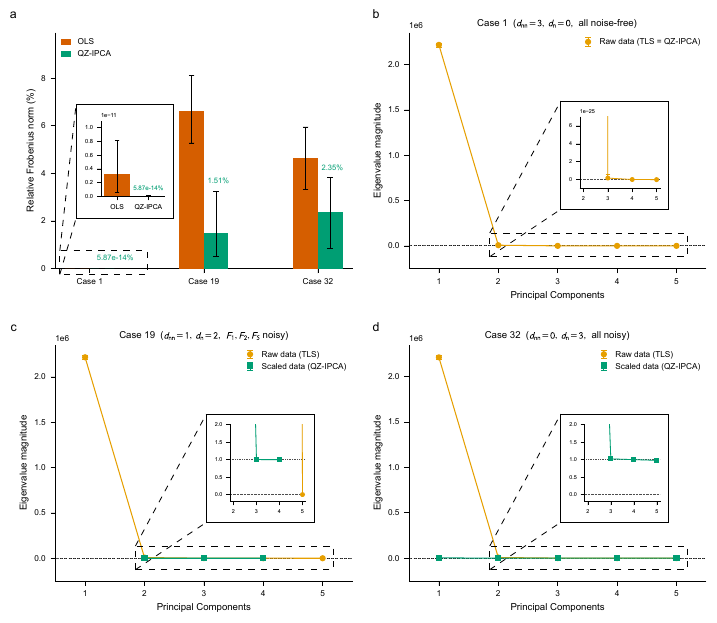}
    \caption{\textbf{OLS and standard TLS fail systematically where QZ-IPCA succeeds, exposing the strength of QZ-IPCA in rank identification as the mechanistic basis for superior coefficient recovery.} (\textbf{a}) Bootstrapped percentage relative Frobenius norm (1000 replicates)  of the estimated regression matrix for OLS (vermillion) and QZ-IPCA (bluish green) across Cases 1, 19, and 32, where the case numbering follows Figure~\ref{fig2:flow_network_cases}. In each replicate, OLS receives the input-output partition identified by QZ-IPCA within that same replicate, that isolates the estimation capability from structural knowledge. Bars are the bootstrap means and error bars correspond to the 95\% confidence intervals. The inset magnifies Case 1. (\textbf{b--d}) Eigenvalue spectra of the raw data covariance matrix (orange circles; the operating objective of TLS under assumed unit noise variances) and of the scaled covariance matrix using QZ-IPCA (green squares; Cases 19 and 32 only) vs. principal component index, where insets magnify components 2--5. Dashed and dotted reference lines mark $y=0$ and $y=1$, the theoretical cluster positions for noise-free (of raw data covariance matrix) and noisy relation eigenvalues (of scaled data covariance matrix), respectively. In \textbf{c}, three finite generalized eigenvalues are sown. The non-noisy dependent variable identified at the PCA stage contributes a zero diagonal in $\mathbf{\Sigma_e}$ of $\mathbf{Sv} = \lambda\mathbf{\Sigma_e v}$, producing one excluded infinite generalized eigenvalue. Panel \textbf{a} establishes that even when granted the best possible input-output partition, OLS cannot recover coefficients with the accuracy of QZ-IPCA; whereas panels \textbf{b}--\textbf{d} reveal that this advantage originates in rank identification, which OLS assumes and standard TLS cannot resolve under heteroskedasticity.}
    \label{fig4:comparison_plot}
\end{figure*}

The OLS comparison was designed to be maximally favorable to the classical method. Since OLS requires a declared input-output partition and cannot infer one from data, we supplied it in each of the 1000 independent bootstrap replicates with the partition identified by QZ-IPCA within that same replicate. This post-hoc consistent partitioning ensures that OLS begins from a robust structural premise in every replicate, so that any remaining performance gap reflects estimation quality alone under the restrictive assumption where the input variables are noise-free. We report the performance of OLS in Figure~\ref{fig4:comparison_plot}\textbf{a}, where OLS yields a relative Frobenius norm of approximately 6.62\% and 4.63\% in Cases 19 and 32, respectively, compared with 1.51\% and 2.35\% for QZ-IPCA. The gap persists even in the noise-free Case 1, where QZ-IPCA achieves $5.87\times10^{-14}\%$, while OLS reaches approximately $3.19\times10^{-12}\%$ (Figure~\ref{fig4:comparison_plot}\textbf{a}, inset). That OLS underperformed even when operating from a robust partition confirms that its limitation is intrinsic to the estimation procedure, i.e., assuming noise-free inputs, and not a consequence of missing structural knowledge. 

The mechanistic basis of this advantage is revealed by the eigenvalue spectra across the same three cases (Figure~\ref{fig4:comparison_plot}\textbf{b}--\textbf{d}). Standard TLS, which under unknown heteroskedastic noise variances reduces operationally to PCA~\cite{Huffel:1991} on the raw data within an implicit unit-variance assumption, succeeds in rank identification only where that assumption is not violated. In Case 1, three near-zero eigenvalues of the raw data covariance matrix unambiguously reveal $d_{\text{non-noisy}}=3$ and both TLS and QZ-IPCA agree (Figure~\ref{fig4:comparison_plot}\textbf{b}). Note that, application of PCA on raw data is itself the first step of our proposed identification algorithm (see Methods), therefore, the same set of eigenvalues represent the results of both TLS and QZ-IPCA methods. Thus QZ-IPCA reaches this conclusion in the initial PCA stage and does not enter the iterative estimation step, as the identification of all five variables in noise-free relation is already complete from the raw spectrum alone. 

\subsection{Statistical consistency and practical operating range}
Until now, the above results were obtained at a single operating point, that is $N=2000$ and $\text{SNR}=10$. We further verified the asymptotic consistency of the noise variances estimated by QZ-IPCA by tracking the mean bias $\mathbb{E}[\hat{\sigma}^2-\sigma^2_{\text{true}}]$ and 95\% CI width across samples sizes $N=50$ to 100,000 on Case 32, in which all five flow variables are noisy. As illustrated in Extended Data Fig. 3, for all five channels, both mean bias and CI width converge monotonically toward zero as $N$ increases, confirming that the variance estimator is asymptotically unbiased and that its variance vanishes in the large-sample limit. Furthermore, we use principal component analysis (PCA) to recover the constraint matrix $(\mathbf{\hat{A}})$ and thereafter the coefficient matrix $(\mathbf{\hat{B}})$ (see Methods) from the scaled data matrix by scaling the raw data matrix using the estimated noise variances. Therefore, as a direct consequence of PCA, the consistency in the variance estimates is an empirical proof of consistency in the coefficient estimates. 

We also stress upon characterizing the full practical operating envelope, and therefore, we evaluated identification success rate of our proposed unified identification algorithm across $N\in\{50,$ $100,$ $200,$ $500,$ $1\text{K},$ $2\text{K},$ $5\text{K},$ $10\text{K}\}$ and $\text{SNR}=\{1,$ $2,$ $5,$ $10,$ $20,$ $50\}$ for all 32 noise configurations (Figure~\ref{fig2:flow_network_cases}\textbf{b}), with 1000 Monte Carlo trials per $(N,\text{SNR})$ grid point. We report the results in Extended Data Fig. 4, where a trial is counted as a success only if rank, non-noisy and noisy variable classification are simultaneously correct. 

\subsection{Robustness to non-Gaussian process excitation}
QZ-IPCA operates on the data covariance matrix, and estimates the measurement noise variances using which it scales the data covariance matrix. The data covariance matrix depends on the true process variables only through their second moment, the method therefore places no distributional assumption on the excitation of the base values of the process variables. This is a property that QZ-IPCA shares with the functional-model estimators~\cite{Gleser:1981} benchmarked here, since OLS, TLS, and GTLS are likewise consistent without requiring the true unobserved regressor values to follow a specified distribution. This distinguishes QZ-IPCA (and the functional-model baselines) from the structural EIV and latent-variable frameworks underlying much of classical multivariate calibration~\cite{Brown:1982} and covariance-structure system identification~\cite{Larimore:1990}, where the true variables are modeled as random draws from a specified distribution; typically Gaussian, chosen for likelihood tractability; rather than as fixed unknowns~\cite{Fuller:1987,Bollen:1989}. To verify that this distribution-independence holds across the full finite-sample operating envelope and not merely asymptotically, we repeated the evaluation under sub-Gaussian (uniform) and super-Gaussian (Laplace) excitation of $F_1$ and $F_2$. The regimes were chosen to bracket excitation statistics seen in real process data, where control-loop nonlinearities such as valve stiction and actuator saturation routinely produce non-Gaussian excitation~\cite{Choudhury:2008}. The excitation variances were held identical to the Gaussian baseline (see Subsection~\ref{subsec:robust-non-gauss} in Methods for more details) with all other simulation aspects unchanged: noise corruption, sample size grid, SNR grid, and success criterion (Extended Data Figs. 5 and 6).

The point-wise deviation in success rate from the Gaussian baseline, $\Delta \text{SR}$ (see Methods), was bounded within $\pm 5.3$ percentage points across all 32 cases and all 48 operating points under both distributional alternatives. Cases 1--16 showed negligible deviation throughout, confirming that excitation distribution is inconsequential when deterministic relations governing identification dominate in number as compared to the stochastic relations. In fully stochastic cases 17--32, deviations were confined to the finite-sample transition zone $N = 100-1000$ with no consistent directional bias, and all converged to within the $\pm3.1\%$ Monte Carlo sampling floor by $N = 2000$. The near-identical worst-case bounds under sub- and super-Gaussian excitation establish that QZ-IPCA is asymptotically distribution-agnostic with respect to the excitation process. This implies that the identification performance in the large-sample limit is insensitive to whether excitation tails are lighter or heavier than Gaussian.

\section{Conclusion} \label{sec:conclusion} 
QZ-IPCA resolves a question that has stood since the classical, EIV, and mixed regression frameworks were formalized separately, that is, whether the structural information conventionally assumed by each (e.g., the noisy/non-noisy partition, noise variances, model rank, and input-output roles) can instead be recovered from data geometry alone. By combining PCA for the identification of exact relations among the non-noisy variables with a QZ-based generalized eigenvalue solver, that remains well-posed under a singular noise covariance, QZ-IPCA subsumes OLS, TLS, and generalized TLS as special cases of a single noise configuration rather than requiring the user to select among them a priori. Validated exhaustively across all $32$ noise topologies of a benchmark flow network, and against classical baselines, the method's advantage traces to rank identification itself rather than to auxiliary structural knowledge.

A natural extension is to dynamic, multi-input multi-output (MIMO) systems, where the constraint relations hold over a time window rather than instantaneously. Integrating the present framework with subspace based dynamic iterative PCA~\cite{Ramnath:2023} and the behavioral approach of system identification~\cite{Willems:1986,Willems:2007,Das:2026} may allow the same partition-free identification to extend to temporal and cross-correlated noise structures. More broadly, the reformulation demonstrated here, that is to let data determine its own structure rather than fixing it by assumption or heuristics, may generalize to other classical estimation problems whose standard solutions rest on comparable prior commitments, including sensor networks and industrial process monitoring, where the boundary between signal and noise is rarely known in advance.




\clearpage
\section{Methods} \label{sec:methods}


\subsection{System description and problem formulation} \label{subsec:ps}
Linear model identification from data separated into three frameworks according to which variables carry measurement noise. In the classical (ordinary least squares, OLS) framework, a set of independent variables is assumed noise-free and a distinct set of dependent variables carries all the noise. So the input-output partition must be declared before estimation~\cite{Kalman:1982,Los:1989}. In the errors-in-variables (EIV, or total least squares, TLS) framework, no such partition exists. Every variable is treated symmetrically and assumed noisy~\cite{Huffel:1991}. In the mixed, or partial-EIV framework, some variables are noise-free and others are not, but which is which must still be supplied by the user~\cite{Huffel:1989}\textsuperscript{,\!}\cite{Huffel:1988}. All three frameworks therefore share a common limitation, which is the noise structure, i.e., which variables are noise, their noise variances, the number of independent relations, and the input-output roles, must be specified a priori, and a misspecification biases the resulting model~\cite{Kalman:1982,Los:1989}.

Formally, consider $n$ variables $(z_1^*,\ldots,z^*_n)$ related by a set of $d$ independent linear relations $f_1,\ldots,f_d \in \mathcal{F}$ satisfying $f_i(z^*_1,\ldots,z^*_n)=0$ for $i=1,\ldots,d$. Given $N$ measurements organized as a data matrix $\mathbf{Z} \in \mathbb{R}^{N\times n}$ (each row one sample, each column one variable with no variable roles or noise labels attached), an offset term is absorbed by augmenting each row with a constant $1$, giving $\mathbf{\tilde{Z}} = \begin{bmatrix} \mathbf{Z} & \mathbf{1} \end{bmatrix} \in \mathbb{R}^{N\times (n+1)}$. The relations are then written in implicit (constraint) form as:
\begin{equation}
    \label{eq:4.1.1}
    \mathbf{A}\left(\mathbf{\tilde{Z}}^*\right)^\top = \mathbf{0}, \qquad \text{subject to}\ \mathbf{Z} = \mathbf{Z}^* + \mathbf{E}
\end{equation}
where $\mathbf{A} \in \mathbb{R}^{d\times(n+1)}$ is the constraint matrix. Unlike explicit regression form $\mathbf{Y}^* =$ $\mathbf{Z}^*_D$ $= \mathbf{Z}^*_I\mathbf{B}^\top$ $= \mathbf{X}^*\mathbf{B}^\top$, Equation~\ref{eq:4.1.1} does not assume which variables are dependent apriori, is unaffected by multicollinearity among the columns of $\mathbf{Z}$, and places no requirement on which variables are noisy. The superscript $(\cdot)^*$ denotes the true unobserved values of the measurements. $\mathbf{A}$ is unique only up to a rotational ambiguity, i.e., left-multiplication by a non-singular matrix. Whereas the true and estimated regression matrices are explicit representations, and therefore they are element-by-element comparable, provided that the choice of dependent (output) and independent (input) variables is consistent during the comparison. Each row of the measurement noise matrix $\mathbf{E}$ is generally assumed to follow a multivariate normal distribution with zero mean and diagonal covariance matrix $\mathbf{\Sigma_e}$, with the assumption that all the noise variances are potentially different, i.e., heteroskedastic. The $n$ noise sequences are further assumed to be uncorrelated with the true underlying unobserved values of the variables for all time instances.

The problem addressed in this work is therefore to estimate, from $\mathbf{\tilde{Z}}$ alone and without any further specification: (i) the partition of the $n$ variables into noisy and non-noisy subsets, (ii) the noise covariance matrix $\mathbf{\Sigma_e}$ of the noisy variables, (iii) the number of linear relations $d$, and (iv) the constraint matrix $\mathbf{A}$ itself. A solution to this problem subsumes OLS, TLS, and partial-EIV or GTLS as special cases of a single noise configuration, rather than requiring the user to select among them in advance.

\subsection{Related work} \label{subsec:related-work}
The prejudice that Kalman identified as intrinsic to noisy identification~\cite{Kalman:1982,Los:1989} takes a different concrete form in each of the frameworks reviewed below, and what varies across them is \emph{which} quantity must be fixed in advance. Two lines of prior work bear directly on this problem.

The first addresses estimation once the noise structure is already known. Classical OLS estimation follows the derivation in \cite[Ch.~14]{ArunSir:2015}. In the EIV setting, the connection between TLS and PCA dates to Pearson~\cite{Pearson:1901}, who used PCA to fit linear relationships among variables all assumed equally noisy, and was placed on a numerical footing by Golub~\cite{Golub:1980} via the SVD of the data matrix. The two formulations are later shown equivalent under homoskedastic noise~\cite[Ch.~8]{Huffel:1991}. The statistical consistency of this estimator, that is, its convergence to the true parameters as the sample size grows, was established for the multivariate EIV regression model by Gleser~\cite{Gleser:1981}. Under heteroskedastic but \emph{known} noise variances $\mathbf{\Sigma_e}$, scaling the data by the Cholesky factor of $\mathbf{\Sigma_e}$ before applying PCA maps zero eigenvalues of the noise-free covariance to unit eigenvalues of the scaled covariance. This procedure directly yields both the relation count $d$ and the constraint matrix $\mathbf{A}$, matching the classical maximum-likelihood results derived for EIV models~\cite{Fuller:1987} and popularized for PCA model identification by Narasimhan and Shah~\cite{ShankarSir:2008}. A more general treatment of known measurement noise, maximum likelihood PCA (MLPCA), accommodates noise covariance that varies from sample to sample and is correlated across both samples and variables, at the cost of of an iterative alternating-regression procedure that neither scales the data directly nor yields a model-order diagnostic as a direct byproduct~\cite{Wentzell:1997}. The scaling-based adopted here and throughout the identification literature that builds on it instead requires only that $\mathbf{\Sigma_e}$ be constant across sample, which is the setting considered in this work. For the mixed case with a declared noisy/non-noisy partition and known variances, GTLS solves the corresponding generalized SVD (GSVD) problem~\cite{Huffel:1989}. The partial-EIV variant, where only a subset of variables is corrupted, admits a dedicated algorithmic treatment once that subset is declared~\cite{Huffel:1988}, with a nonlinear adjustment reformulation given by Xu et al.~\cite{Xu:2012} In every one of these formulations, the partition, the variances, or both, enter as inputs rather than outputs of the estimation procedure.

The second line addresses the case where $\mathbf{\Sigma_e}$ is unknown but all variables are noisy, relaxing the requirement that variances be supplied while still assuming every variable is corrupted. Estimating a constraint matrix via PCA has itself been framed as a data reconciliation problem~\cite{ShankarSir:2015}, a perspective extended by iterative PCA (IPCA)~\cite{ShankarSir:2008}. Simultaneously estimating both the model and the noise covariance by unconstrained maximum likelihood is not merely difficult but ill-posed. Even in the univariate functional-model case, the join-likelihood is unbounded and admits no consistent solution~\cite{Fuller:1987}, so a constrained modification of the likelihood equation is required to obtain a consistent estimator~\cite{Chan:1983}. IPCA implements such a modification by alternating between them: given an estimate of noise covariance $\mathbf{\hat{\Sigma}_e}$ , scaled PCA yields $\mathbf{\hat{A}}$; given $\mathbf{\hat{A}}$, the constraint residuals $\mathbf{r}_i = \mathbf{\hat{A}} \mathbf{\tilde{z}}_i$ yield a maximum-likelihood update of $\mathbf{\hat{\Sigma}_e}$ by minimizing:
\begin{equation}
    \label{eq:4.2.1}
    \min_{\mathbf{\Sigma_e}} N\log\left| \mathbf{\hat{A}}\mathbf{\Sigma_e}\mathbf{\hat{A}}^\top \right| + \sum_{i=1}^N \mathbf{r}_i^\top \left( \mathbf{\hat{A}} \mathbf{\Sigma_e} \mathbf{\hat{A}}^\top \right)^{-1} \mathbf{r}_i
\end{equation}
subject to the identifiability condition $d(d+1)/2 \geq n$. This bound counts the free parameters in $\mathbf{\Sigma_e}$ against the number of independent equations involved in the constraint-residual covariance computed as $\mathbf{\Sigma_r} = \mathbf{\hat{A}}\mathbf{\hat{\Sigma}_e}\mathbf{\hat{A}}^\top$, that are available to estimate the free parameters. Assuming $\mathbf{\Sigma_e}$ to be diagonal, as adopted throughout this work, reduces the number of free parameters to $n$ and yields exactly the bound above. A non-diagonal but sparse $\mathbf{\Sigma_e}$ can in principle be accommodated by the same counting argument, but only if the locations of the nonzero off-diagonal entries are specified in advance. This can be considered as a lighter prior commitment than declaring the full partition or all variances, but a prior commitment nonetheless. IPCA is confined to the strict EIV case. However, any non-noisy variable makes $\mathbf{\Sigma_e}$ singular and Equation~\ref{eq:4.2.1} ill-posed, so the method cannot be applied at all once even one variable is exact. It is not merely less accurate under mixed noise, it is undefined. 

No existing method, then, estimates the noisy/non-noisy partition, $\mathbf{\Sigma_e}$, $d$, and $\mathbf{A}$ simultaneously from unlabeled data. Each of the frameworks above resolves this ambiguity Kalman described by fixing part of it externally, either the partition (OLS, GTLS, partial-EIV), the relative noise scale (TLS), or the requirement that every variable be noisy at all (IPCA). Closing this gap, without reintroducing any of these prior commitments, is the contribution of the framework described below.

\subsection{Unifying identification framework}
Model identification begins by separating the $d$ linear independent relations into those involving only non-noisy variables (deterministic relations) and those involving at least one noise variable (stochastic relations), denoted as $d_{\text{non-noisy}}$ and $d_{\text{noisy}}$, respectively. However, this split is not always visible by inspection. For instance, consider the following set of equations:
\begin{subequations}
    \label{eq:4.3.1}
    \begin{align}
        z_3^* &= z_1^* + z_2^*, \label{eq:4.3.1a} \\
        z_3^* &= z_4^*, \label{eq:4.3.1b} \\
        z_5^* &= z_4^* - z_2^* \label{eq:4.3.1c}
    \end{align}
\end{subequations}
Now, suppose the given measurements of all the five variables contain variable $z_1$ and $z_5$ as exact, i.e., non-noisy variables. Even though initially, it is not explicitly visible that any constraint exists solely between the non-noisy variables, a linear combination of the above relations that individually mix noisy and non-noisy variables yields the relation $z_5^* = z_1^*$. 

Such exact relations manifest as zero eigenvalues of the raw data covariance matrix, which motivates the first of four concepts underlying the framework. As an outline of the complete identification framework, reader is referred to see Extended Data Fig.~1.

\noindent \textbf{Concept 1: Identification of non-noisy relations}. Let $\boldsymbol{\mathcal{S}} = \mathbf{\tilde{Z}}^\top \mathbf{\tilde{Z}}/N$ be the raw data covariance matrix. Relations involving only non-noisy variables, $\mathbf{A}_{\text{non-noisy}}$, correspond exactly to the eigenvectors of the zero eigenvalues of $\boldsymbol{\mathcal{S}}$. This provides the number of such constraints, $d_{\text{non-noisy}}$, and the non-zero entries of these eigenvectors identify the first subset of non-noisy variables, represented as $n_{\text{non-noisy}}^{(1)}$. Further non-noisy variables may remain undetected at this stage because of potentially being a part of the noisy (stochastic) relations. Therefore, the total number of non-noisy variables can be considered as the sum of two of its subset, $n_{\text{non-noisy}} = n_{\text{non-noisy}}^{(1)} + n_{\text{non-noisy}}^{(2)}$, identified at different stages. At this stage, only the first subset is identified, and the remaining variables comprise the noisy variables, $n_{\text{noisy}}$ and the second subset of non-noisy variables $n_{\text{non-noisy}}^{(2)}$, that are yet to be determined. 

The subset $n_{\text{non-noisy}}^{(1)}$ is then partitioned into independent and dependent variables (see Supplementary Note 1, which follows the procedure of Narasimhan \& Bhatt~\cite{ShankarSir:2015}), and the $d_{\text{non-noisy}}$ dependent variables among them, being linearly redundant with the independent variables, are removed from the data matrix $\mathbf{\tilde{Z}}$. This leaves a reduced data matrix $\mathbf{Z}_{N\times p}$ with $p$ comprising: 
\begin{equation}
    \label{eq:4.3.2}
    p = \left(n^{(1)}_{\text{non-noisy}} - d_{\text{non-noisy}}\right) + n^{(2)}_{\text{non-noisy}} + n_{\text{noisy}} + 1
\end{equation}
with the $+1$ accounting for the offset column. Because $\mathbf{Z}_{N\times p}$ may still contain non-noisy variables, its noise covariance $\mathbf{\Sigma_e}$ is potentially singular, and the scaling used by IPCA is not directly applicable. This leads to the second concept. \\

\noindent \textbf{Concept 2: Identification of constraints via generalized eigendecomposition.} The remaining constraints are instead obtained from the generalized eigenvectors $\mathbf{v}$ of the covariance matrix $\mathbf{S}$ of $\mathbf{Z}_{N\times p}$ with respect to the corresponding noise covariance matrix $\mathbf{\Sigma_e}$ by solving the following generalized eigenvalue problem:
\begin{equation}
    \label{eq:4.3.3}
    \mathbf{S} \mathbf{v} = \lambda\ \mathbf{\Sigma_e} \mathbf{v}
\end{equation}
by employing the QZ algorithm~\cite{Moler:1973}. The equivalence between the scaled PCA formulation and the generalized eigenvalue problem arising here is formally demonstrated in Supplementary Note 2. The QZ algorithm remains well-defined when $\mathbf{\Sigma_e}$ is singular, where the zero diagonal entries of $\mathbf{\Sigma_e}$ (corresponding to the non-noisy variables) map to infinite generalized eigenvalues rather than causing failure (see Supplementary Note 3). This equivalently solves the GTLS problem~\cite{Huffel:1989} without requiring $\mathbf{\Sigma_e}$ or the partition to be supplied. These remaining constraints, denoted as $\mathbf{A}_{\text{noisy}}$, are subsequently derived using the third concept. \\
 
\noindent \textbf{Concept 3: Identification of noisy-relation subspace.} Considering involving noisy variables, $\mathbf{A}_{\text{noisy}}$ (which may also involve non-noisy variables), correspond to the eigenvectors of the smallest \emph{equal} finite generalized eigenvalues of Equation~\ref{eq:4.3.3}. Under correct variance scaling, these eigenvalues cluster at unity, giving $d_{\text{noisy}}$ directly. Kindly refer to the Supplementary Note 4, where we provide the formal hypothesis test that we employ to estimate $d_{\text{noisy}}$. \\

\noindent \textbf{Concept 4: Iterative estimation of the noise variances.} $\mathbf{\Sigma_e}$ is estimated for the variables retained in $\mathbf{A}_{\text{noisy}}$ by the QZ-based IPCA optimization by integrating Equation~\ref{eq:4.2.1} with the QZ algorithm (see Extended Data Fig.~2), iterated jointly with Equation~\ref{eq:4.3.3} until $\lambda$ converges and the smallest $d$ eigenvalues test as statistically equal (we employ hypothesis test by following the work of Jolliffe~\cite{Jolliffe:2002}. See Supporting Note 4 for more details). This step may itself reveal further non-noisy variables $(\sigma^2\to 0)$, which are grouped into $n^{(2)}_{\text{non-noisy}}$. The degrees of freedom are search from $d_{\max}=p-1$ down to the minimum permitted value obtained by the identifiability condition:
\begin{equation}
    \label{eq:4.3.4}
    \frac{d_{\text{noisy}}(d_{\text{noisy}}+1)}{2} \geq \left\{ p - \left( n_{\text{non-noisy}}^{(1)} - d_{\text{non-noisy}}\right) -1\right\}
\end{equation}

Combining Concepts 1--4 (see Extended Data Fig.~1) yields the full constraint matrix $\mathbf{A} = \begin{bmatrix} \mathbf{A}_{\text{non-noisy}}^\top & \mathbf{A}_{\text{noisy}}^\top \end{bmatrix}^\top$, without requiring any prior declaration of variable roles, noise status, variances, or model order. 

In the degenerate case $n^{(1)}_{\text{non-noisy}}=n$ (all variables resolved at the PCA stage), the procedure terminates with $\mathbf{A} = \mathbf{A}_{\text{non-noisy}}$, $d = d_{\text{non-noisy}}$, $\mathbf{\Sigma_e} = \mathbf{0}$, recovering the noise-free case directly. Because the offset column of $\mathbf{\tilde{Z}}$ is appended as an exact constant, its entry in $\mathbf{\Sigma_e}$ is fixed at zero, so the offset is estimated without measurement noise by construction.

\subsection{Benchmark system and noise topology enumeration}
All the simulation experiments were conducted on a five-variable flow network comprising three process units and five flow streams ($F_1$--$F_5$) as presented in Figure~\ref{fig2:flow_network_cases}\textbf{a}, originally introduced by the IPCA authors~\cite{ShankarSir:2008}. This network is governed by three linear mass balance relations under the steady-state operating condition where the flow rates remain constant over time, with the true constrained model $\mathbf{A}_{\text{true}}$ relating the flow variables $\mathbf{F} = \begin{bmatrix} F_1 & F_2 & F_3 & F_4 & F_5 \end{bmatrix}^\top$ as:
\begin{equation}
    \label{eq:4.4.1}
    \mathbf{A}_{\text{true}}\mathbf{F} = \begin{bsmallmatrix} 1 & 1 & -1 & 0 & 0 \\ 0 & 0 & 1 & -1 & 0 \\ 0 & -1 & 0 & 1 & -1 \end{bsmallmatrix} \mathbf{F} = \mathbf{0}
\end{equation}
where the rank of $\mathbf{A}_{\text{true}}$, i.e., $d = d_{\text{non-noisy}} + d_{\text{noisy}} = 3$, gives the total relation dimension. $d_{\text{non-noisy}}$ and $d_{\text{noisy}}$ (often denoted as $d_{nn}$ and $d_{n}$ in this work, respectively) denote the number of noise-free and noisy linear relations, respectively, and the precise partition between them is determined by the noise topology of each case. 

The complete combinatorial space of noise assignments across the five binary-classified variables yields $2^5=32$ distinct noise configurations. These are enumerated in canonical order and grouped by the number of noisy variables (0 through 5), as illustrated in Figure~\ref{fig2:flow_network_cases}\textbf{b}, with case numbering consistent across all figures and tables in this work. In all 32 configurations, QZ-IPCA receives only the raw data matrix with no specification of noise status, noise variances, variable roles, or relation count.

\subsection{Data generation} \label{subsec:data-gen}
The two flow variables $F_1$ and $F_2$ were considered to be free variables each with a base operating point value of $10$ L per minute. True (noise-free) flow values (distinguished by an asterisk in the superscript) of these two variables were generated by adding normally distributed random fluctuations with zero-mean and excitation variances $\sigma^2_{F_1^*}=1$ and $\sigma^2_{F_2^*}=4$, respectively, to their respective base operating point values. The true values of the remaining flow variables $F_3,F_4,$ and $F_5$ were computed deterministically from the linear flow balance constraints (Equation~\ref{eq:4.4.1}) governing the network. The signal variances of each channel, $\sigma^2_{F_i^*}$, was computed empirically from the generated true data matrix and held fixed for all the simulation runs at a specific samples size $N$. 

For each noise configuration, zero-mean Gaussian measurement noise was added independently to the variables designated as noisy in that configuration. The noise variance for noisy variable $F_i$ was set as:
\begin{equation}
    \label{eq:4.5.1}
    \sigma^2_{i} = \sigma^2_{F_i^*} \big/ \mathrm{SNR}
\end{equation}
which was computed once from the true data and held fixed across all replicate runs at that operating point (i.e., the number of samples $N$ and the signal-to-noise ratio SNR). This parameterization ensures that the SNR is defined consistently with respect to the actual signal power of each channel. Variables designated as noise-free received no additive noise. At the baseline operating condition of $N=2000$ and $\text{SNR}=10$, the resulting true per-channel noise variances are as follows: 
\begin{equation}
    \label{eq:4.5.2}
    \sigma_{F_1}^2 = 0.108, \quad \sigma_{F_2}^2 = 0.399, \quad \sigma_{F_3}^2 = 0.503, \quad \sigma_{F_4}^2 = 0.503, \quad \sigma_{F_5}^2 = 0.108
\end{equation}
These values are never disclosed to the method and serve solely as the ground truth against which estimates were evaluated. Due to the structural symmetry of the network, $\sigma^2_{F_1} = \sigma^2_{F_5}$ and $\sigma^2_{F_3}=\sigma^2_{F_4}$, while $\sigma^2_{F_2}$ is distinct from both pairs, yielding three distinct variance levels within the same system (heteroskedasticity). All noise sequences were drawn independently across channels, across time steps, and across replicate runs, and are statistically independent of the corresponding true signal values, consistent with the model assumptions stated in Subsection~\ref{subsec:ps}. 


\subsection{Bootstrap simulation protocol}
While characterizing the performance across all 32 noise configurations (Figure~\ref{fig3:32_cases_results}, Extended Data Tables 1--3) and analyzing the comparative benchmarking (Figure~\ref{fig4:comparison_plot}), the baseline operating condition was used, i.e., $N=2000$ and $\text{SNR}=10$. At each noise configuration, 1000 independent simulation runs were carried out, each with a freshly generated data matrix of $N$ observations by following the data generation procedure described above. We applied our identification algorithm independently to each replicate, yielding replicate-specific estimates of the noise variances $\hat{\sigma}^2_i$, number of non-noisy and noisy relations (the relation dimension), constraint matrix $\mathbf{\hat{A}}$, variable partition, and consequently the coefficient matrix $\mathbf{\hat{B}}$, i.e., the regression matrix.

All reported point estimates, e.g., mean variance estimates, mean normalized Frobenius errors, mean maximum element-wise absolute error, and mean relation dimension estimates, represent the empirical mean across the 1000 replicates. All reported 95\% confidence intervals were constructed by the non-parametric percentile method, taking 2.5th and 97.5th percentiles of the empirical bootstrap distribution across 1000 replicate estimates. This procedure makes no distributional assumption about the estimator and is exact for any continuous sampling distribution in the large replicate limit. Confidence intervals are reported throughout Extended Data Tables 1--2 in the format [Lower Bound, Upper Bound]. 

\subsection{Performance metrics}
Four complementary metrics were used to characterize QZ-IPCA performance across the 32 configurations. 
\begin{itemize}
    \item First, rank identification was assessed as a binary outcome for each of $d_{\text{non-noisy}}$ and $d_{\text{noisy}}$ separately. A configuration is classified as correct if both estimated relation dimensions match the true values exactly. 
    \item Second, absolute variance deviation $\left|\hat{\sigma}^2_i - \sigma^2_{\text{true},i}\right|$ was computed per channel for all variables identified as noisy, providing a direct measure of the accuracy of the noise characterization step. 
    \item Third, the normalized Frobenius coefficient error was computed as:
        \begin{equation}
            \label{eq:4.7.1}
            \delta_F(\%) = \left( \big\| \mathbf{B}_{\text{true}} - \mathbf{\hat{B}} \big\|_F\ \Big/\ \big\| \mathbf{B}_{\text{true}} \big\|_F\right) \times 100
        \end{equation}
        where $\|\cdot\|_F$ denotes the Frobenius norm, quantifying the overall accuracy of the recovered regression coefficient matrix relative to its true magnitude. We remained consistent with the use of the same partition of possible set of independent and dependent variables, produced by our algorithm, to compute both $\mathbf{B}_{\text{true}}$ and $\mathbf{\hat{B}}$ from the true and estimated constrained matrices $\mathbf{A}_{\text{true}}$ and $\mathbf{\hat{A}}$, respectively, in each bootstrap simulation trials. Note that, the partition may vary across different trials due to the condition number of the possible dependent variable set, based on which the partitioning took place.
    \item Fourth, the maximum element-wise absolute error was computed as:
        \begin{equation}
            \label{eq:4.7.2}
            \big\| \mathbf{B}_{\text{true}} - \mathbf{\hat{B}} \big\|_{\infty} = \underset{i,j}{\max}\ \big| B_{\text{true}\ i,j} - \hat{B}_{i,j} \big|
        \end{equation}
        which is a measure of the supplementary worst-case coefficient recovery, reported alongside the offset term error $\|\mathbf{\hat{b}}\|_\infty$ in Extended Data Table 3. Across 32 cases analyzed, the 95\% confidence intervals for every individual offset term spanned zero (not reported in the table), confirming that the recovered relations are statistically unbiased with respect to the zero constant term.
\end{itemize}

\subsection{Comparative benchmarking against OLS and TLS}
Comparative analysis was performed on three cases selected to span the noise-topology space: Case 1 ($d_{\text{non-noisy}}=3,$ $d_{\text{noisy}}=0$; all variables are noise-free), Case 19 ($d_{\text{non-noisy}}=1,$ $d_{\text{noisy}}=2$; $F_1,$ $F_2,$ $F_5$ are noisy), and Case 32 ($d_{\text{non-noisy}}=0,$ $d_{\text{noisy}}=3$; all variables are noisy). Case 19 was selected as a representative 3-noisy case with non-zero deterministic and stochastic model ranks; the structural argument for why it is representative follows from the group-level properties, not from anything particular to that specific case number. These cases were evaluated using the same 1000-replicate bootstrap protocol at $N=2000$ and $\text{SNR}=10$. \\

\noindent \textbf{OLS comparison}. OLS requires an explicit input-output partition of the variables before estimation can proceed, a quantity QZ-IPCA does not require and the comparison study does not provide externally. To construct a maximally fair comparison, OLS was supplied in each replicate with the input-output partition identified by QZ-IPCA within that same replicate. The procedure, i.e., post-hoc consistent partitioning, ensures that OLS begins from a robust structural premise in every replicate and that any remaining performance gap reflects estimation quality rather structural ignorance. In each replicate, both OLS and QZ-IPCA estimate their respective regression matrices under an identical partition, and the percentage relative Frobenius norm $\delta_F$ (Equation~\ref{eq:4.7.1}) is computed for both. The partition supplied to OLS may differ across replicates, as QZ-IPCA's identification is itself subject to sampling variability. Therefore the bootstrap distribution of $\delta_F$ for OLS incorporates both estimation variance and replicate-to-replicate partition variation. \\
    
\noindent \textbf{TLS comparison and GTLS exclusion}. Standard TLS, under the assumption of homoskedastic noise with unknown equal variances, treats all variables symmetrically. When noise variances are unknown, which is the operating condition of this study, TLS cannot perform any variance-informed rescaling and reduces operationally to PCA of the raw data covariance matrix with implicit unit-variance assumptions. The rank-determining object for TLS is therefore the raw data eigenvalue spectrum, and the number of linear relations is inferred as the number of smallest eigenvalues that are equal to each other. GTLS~\cite{Huffel:1989} was excluded from the comparison because it requires both the noise variances and the non-noisy/noisy variable partition as explicit inputs. In the absence of these quantities, GTLS cannot be executed and is operationally indistinguishable from standard TLS. Including it would therefore require supplying it with the ground-truth values it was designed to assume, which would render the comparison uninformative.
    
For the eigenvalue comparison in Figure~\ref{fig4:comparison_plot}\textbf{b}--\textbf{d}, two types of eigenvalues are displayed per case where applicable: the eigenvalues of the raw data covariance matrix, which constitutes the TLS operating object; and the eigenvalues of the iteratively variance-scaled covariance matrix produced by QZ-IPCA after the convergence of the generalized eigenvalue iteration (Extended Data Figs. 1 and 2). Reference lines at $y=0$ and $y=1$ mark the theoretical cluster positions for eigenvalues corresponding to noise-free and noisy linear relations, respectively, under correct variance estimation. In Case 19, only three finite generalized eigenvalues of the scaled data covariance matrix are displayed. Upon identifying the non-noisy dependent variable at the initial PCA stage (Extended Data Fig. 1), QZ-IPCA removes that variable from the iterative estimation step (Extended Data Fig. 2). The diagonal entry corresponding to the non-noisy independent variable in the noise covariance matrix $\mathbf{\Sigma_e}$ in the generalized eigen-problem $\mathbf{Sv} = \lambda\mathbf{\Sigma_e v}$ is therefore zero, producing one infinite generalized eigenvalue that is excluded from the display in Figure~\ref{fig4:comparison_plot}\textbf{c}.

\subsection{Asymptotic consistency analysis}
The asymptotic behavior of our proposed QZ-IPCA's variance estimator was analyzed by considering Case 32 in which all five variables carry measurement noise and the fully iterative estimation procedure is always invoked. The sample sizes considered to conduct the experiment were $N\in\{50,$ $100,$ $200,$ $500,$ $1\text{K},$ $2\text{K},$ $5\text{K},$ $10\text{K},$ $50\text{K},$ $100\text{K}\}$, with SNR held fixed at 10. At each value of $N$, 1000 independent bootstrap simulation runs were performed following the data generation procedure (Subsection~\ref{subsec:data-gen}). The bias in the estimated noise variance value for channel $i$ was computed as the empirical mean residual:
\begin{equation}
    \label{eq:4.8.1}
    \text{Bias}(N) = \frac{1}{1000} \times \sum_{k=1}^{1000} \left( \hat{\sigma}^2_{i,k} - \sigma^2_{\text{true},i} \right) = \mathbb{E}\left[ \hat{\sigma}^2_{i,k} - \sigma^2_{\text{true},i} \right]
\end{equation}
where $\hat{\sigma}^2_{i,k}$ is the estimate from the $k$-th run. The 95\% confidence interval at each $N$ was constructed by the percentile method across 1000 residuals $(\hat{\sigma}^2_{i,k} - \sigma^2_{\text{true},i})$. Results for all five channels individually and combined are reported in Extended Data Fig. 3. 
%

\subsection{Operating envelope evaluation} \label{subsec:operating-envelope}
The practical operating range of QZ-IPCA wad characterized across all 32 noise configurations simultaneously over a grid of sample sizes $N=\{50,$ $100,$ $200,$ $500,$ $1\text{K},$ $2\text{K},$ $5\text{K},$ $10\text{K}\}$ and $\text{SNR}\in\{1,2,5,10,20,50\}$, reported in Extended Data Fig. 4. Panels are ordered consistently with Figure~\ref{fig2:flow_network_cases}\textbf{b}. At each operating point $(N, \text{SNR})$, 1000 independent Monte Carlo simulation trials were performed in order to compute success percentage of our proposed identification method. In each trial, a fresh data matrix of $N$ observations was generated according to the procedure described in Subsection~\ref{subsec:data-gen}, with noise variance for each noisy channel set as defined in Equation~\ref{eq:4.5.1}, computed once from the true data matrix and held fixed across all trials at that operating point.

A trial was recorded as a success only if all three of the following conditions were simultaneously satisfied: (i) the number of noise-free relations $d_{\text{non-noisy}}$ was correctly identified at the initial PCA stage; (ii) the number of noisy relations $d_{\text{noisy}}$ was correctly identified by the QZ-IPCA iterative stage; and (iii) the complete noisy and non-noisy variable partition was correct. Apart from the variables in the set $n^{(1)}_{\text{non-noisy}}$; i.e., those identified deterministically by PCA at the first stage; a variable was classified as noise-free if its estimated variance satisfied:
\begin{equation}
    \label{eq:4.10.1}
    \hat{\sigma}^2 < \left(0.1\times \underset{i:\sigma^2_{i}>0}{\min}\ \sigma^2_{i}\right)
\end{equation}
where the minimum is taken over all truly noisy variables in that configuration. The success rate at each operating point carries a 95\% confidence half-width, computed as:
\begin{equation}
    \label{eq:4.10.2}
    \text{CI half width} = \pm1.96\times \sqrt{\frac{p(1-p)}{\#\ \text{MC trials}}}
\end{equation}
which turns out to be $\pm1.96\times\sqrt{0.25/10^3}$ $\simeq$ $\pm3.1\%$, derived from the normal approximation to the binomial distribution at the conservative maximum-variance point of $p=0.5$. Operating points at which success rate deviations from a reference do no exceed $\pm0.031$ should be interpreted as consistent with that reference within Monte Carlo sampling uncertainty. 

\subsection{Robustness to non-Gaussian excitation} \label{subsec:robust-non-gauss}
To verify that the performance of our proposed identification algorithm is insensitive to the distributional form of the process excitation, the complete operating-envelope evaluation was repeated under two non-Gaussian excitation regimes, using the identical $(N,\text{SNR})$ grid, 32 noise configurations, 1000 Monte Carlo trials per operating points, as success criterion as described in Subsection~\ref{subsec:operating-envelope}. All aspects of data generation for the dependent variables, corruption of true values with noise sequences were unchanged; only the distribution from which the free variables $F_1$ and $F_2$ were excited was altered.

\begin{itemize}
    \item \textbf{Sub-Gaussian excitation}. Free variables were excited using independent zero-mean uniform distribution on $[-\sqrt{3V_i},\ +\sqrt{3V_i}]$, where $V_i = \sigma^2_{F_i^*}$ is the excitation variance of channel $i=\{1,2\}$, held identical to the Gaussian baseline (refer to Subsection~\ref{subsec:data-gen}). This parameterization preserves the excitation variance exactly while reducing the kurtosis to 1.8, below the Gaussian value of 3. 
    \item \textbf{Super-Gaussian excitation}. Free variables were using independent zero-mean Laplace distribution with scale parameter $b_i=\sqrt{V_i/2}$, where $V_i=\sigma^2_{F_i^*}$ as above. This parameterization again preserves the excitation variance while introducing heavier tails, with excess of 3 relative to the Gaussian baseline. 
\end{itemize}

Performance under each non-Gaussian regime was quantified by the point-wise success rate deviation:
\begin{equation}
    \label{eq:4.11.1}
    \Delta \text{SR} = \text{SR}_{\text{non-Gaussian}} - \text{SR}_{\text{Gaussian}}
\end{equation}
where $\text{SR}_{\text{Gaussian}}$ is taken directly from Extended Data Fig. 4 as the baseline. Results for sub-Gaussian and super-Gaussian excitation are reported separately in Extended Data Figs. 5 and 6, respectively, with identical colormap range and axis conventions to permit direct visual comparison across both non-Gaussian regimes and between each regime and the Gaussian baseline. Operating points with $|\Delta \text{SR}| \leq 0.031$ are indistinguishable from zero within Monte Carlo sampling noise and are interpreted as consistent with the Gaussian baseline. 

%



\subsection*{Data availability} \label{sec:data-avail}
In order to apply our proposed algorithm, the data were generated by following the guidelines presented in Subsection~\ref{subsec:data-gen}, thus no external datasets were used. \\



\noindent \textbf{Author contributions} M.S.K.G. designed and implemented the algorithm. M.S.K.G. and D.D. conducted the experiments. D.D. created the figures and wrote the initial draft. A.K.T. and S.N. jointly supervised the work, suggested experiments, contributed ideas, and provided extensive feedback. \\

\noindent \textbf{Competing interests} The authors declare no competing interests.

\subsection*{Additional information}
\textbf{Supplementary Information}: A separate supplementary PDF document containing extended methods is available as an ancillary file on the arXiv abstract page for this preprint. \\ 

\noindent \textbf{Correspondence and requests for materials} should be addressed to Arun K. Tangirala and Shankar Narasimhan.

%
%

\clearpage
\begin{figure*}[ht]
  \centering
  \includegraphics[width=89mm]{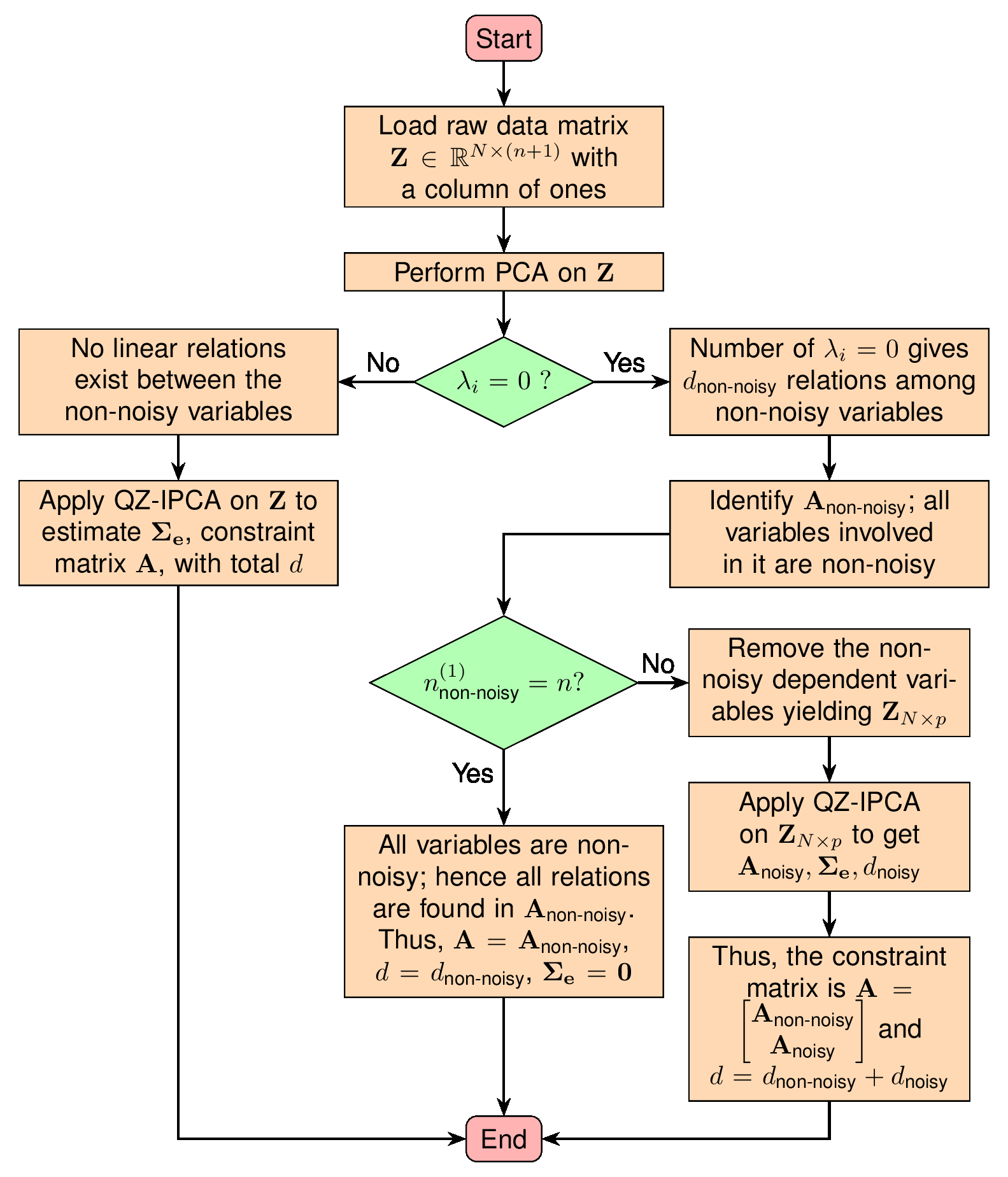} 
  \caption{\textbf{Extended Data Fig. 1 $\mid$ A unified algorithmic pipeline for autonomous model estimation.} The framework requires only the raw unlabeled data matrix ($\mathbf{Z}$) as input, bypassing the need for prior assumptions regarding noise structure or variable partitioning. By sequentially isolating non-noisy variables via PCA, the algorithm adaptively applies QZ-IPCA (detailed in Extended Data Fig. 2) to extract the constraint matrix ($\mathbf{A}$), the exact number of underlying linear relations ($d$), and the noise covariance ($\mathbf{\Sigma_e}$). This generalized execution naturally subsumes classical methods; including ordinary, total, and generalized least squares; into a single, assumption-free procedure capable of resolving any noise configuration.}
\end{figure*}

\clearpage
\begin{figure*}[ht]
  \centering
  \includegraphics[width=89mm]{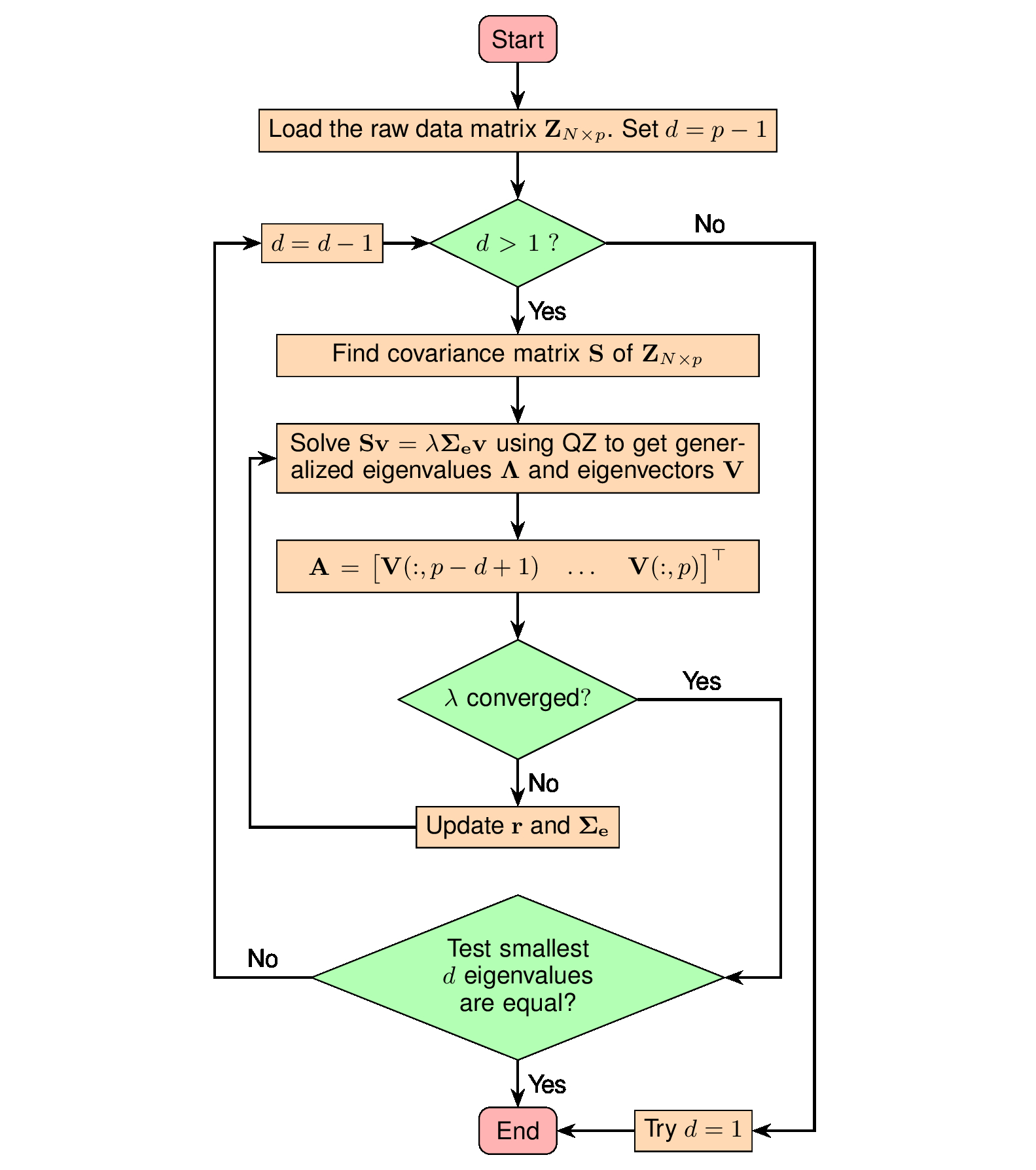} 
  \caption{\textbf{Extended Data Fig. 2 $\mid$ Iterative generalized eigenvalue solver for QZ-IPCA.} The core computational engine of the estimation framework. The algorithm dynamically resolves the generalized eigenvalue problem $\mathbf{Sv} = \lambda \mathbf{\Sigma_e v}$ utilizing the QZ algorithm~\protect\cite{Moler:1973}. Iterating through potential degrees of freedom ($d$), starting from a maximum possible value $d_{\max}=p-1$ to $d_{\min}$ constrained by the identifiability condition given in Equation~\ref{eq:4.3.4}, the procedure continuously updates the estimated noise covariance ($\mathbf{\Sigma_e}$) until convergence is achieved, that is verified using a hypothesis test for the equality of the smallest $d$ eigenvalues. This mathematically guarantees the optimal recovery of the generalized eigenvectors $\mathbf{v}$ used to construct the final constraint matrix $\mathbf{A}$ directly from a potential mixture of noisy and noise-free data.}
\end{figure*}

\clearpage
\begin{figure*}[ht]
  \centering
  \includegraphics[width=\textwidth]{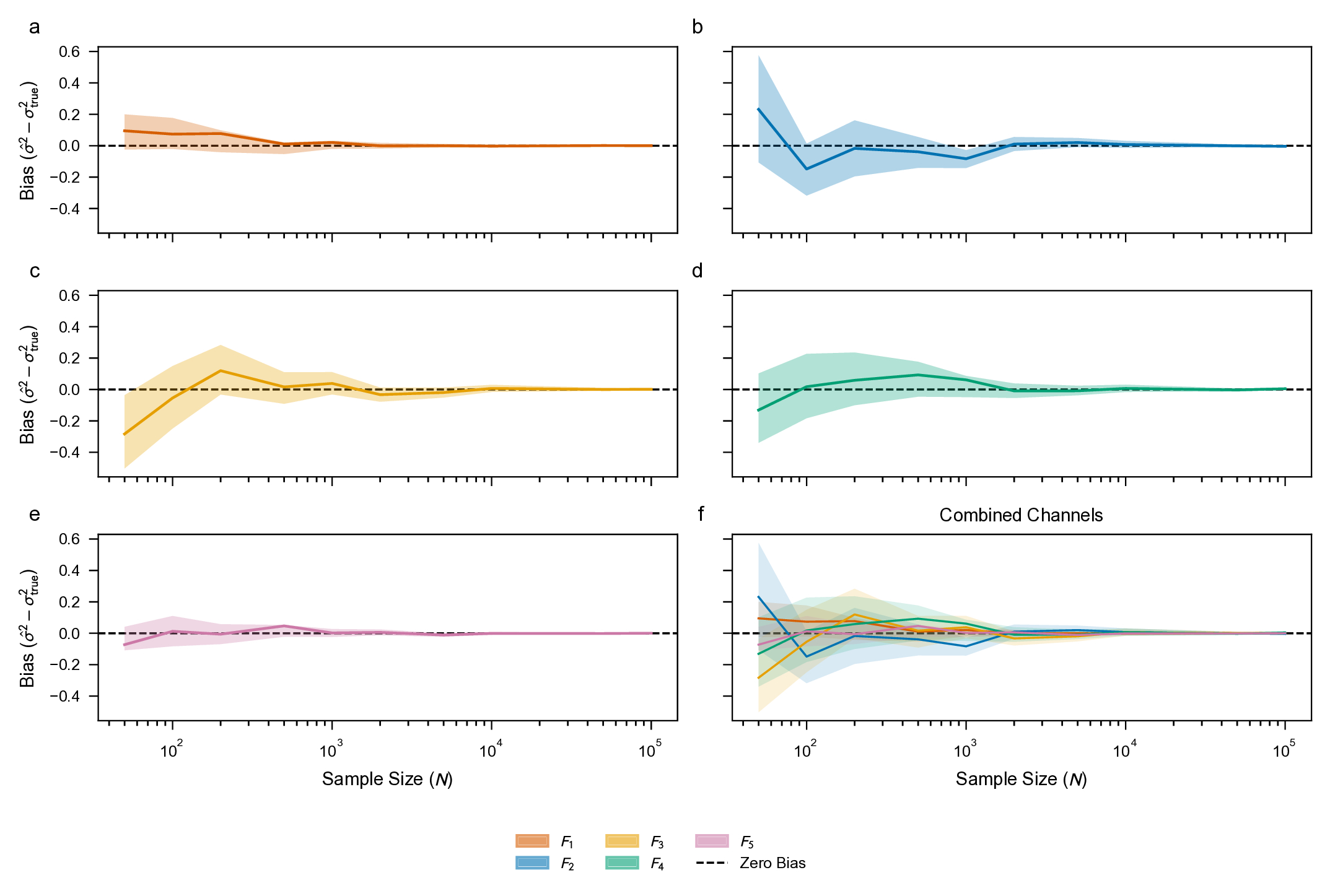}
  \caption{\textbf{Extended Data Fig. 3 $\mid$ Asymptotic consistency of QZ-IPCA variance and coefficient estimates.} The above analysis considers the 32nd noise topology, where all the flow variables are noisy. Panels (\textbf{a})--(\textbf{e}) show the bias in noise variance estimation for each flow variable ($F_1$--$F_5$) individually as a function of sample size $N$ (SNR = 10 across all channels). Panel (\textbf{f}) overlays all five channels for direct comparison. Solid lines denote the mean bias $\mathbb{E}\left[ \hat{\sigma}^2  - \sigma^2_{\text{true}}\right]$, estimated from 1000 independent bootstrap simulation runs at each $N$; shaded regions define the 95\% confidence intervals computed via the percentile method across the same 1000 residuals $\left( \hat{\sigma}^2 - \sigma^2_{\text{true}} \right)$. In all channels, both the mean bias and the CI width converge monotonically toward zero as $N$ increases, confirming that the variance estimator is asymptotically unbiased and that its variance vanishes in the large-sample limit; which are the two conditions that together constitute statistical consistency. Furthermore, the constrained matrix is retrieved from the scaled data matrix by applying PCA, where scaling factors are the estimated noise variances. Therefore, consistency in the variance estimates results in the consistency of the model coefficient estimates as a direct consequence of PCA.}
\end{figure*}

\clearpage
\begin{figure*}[ht]
  \centering
  \includegraphics[width=\textwidth]{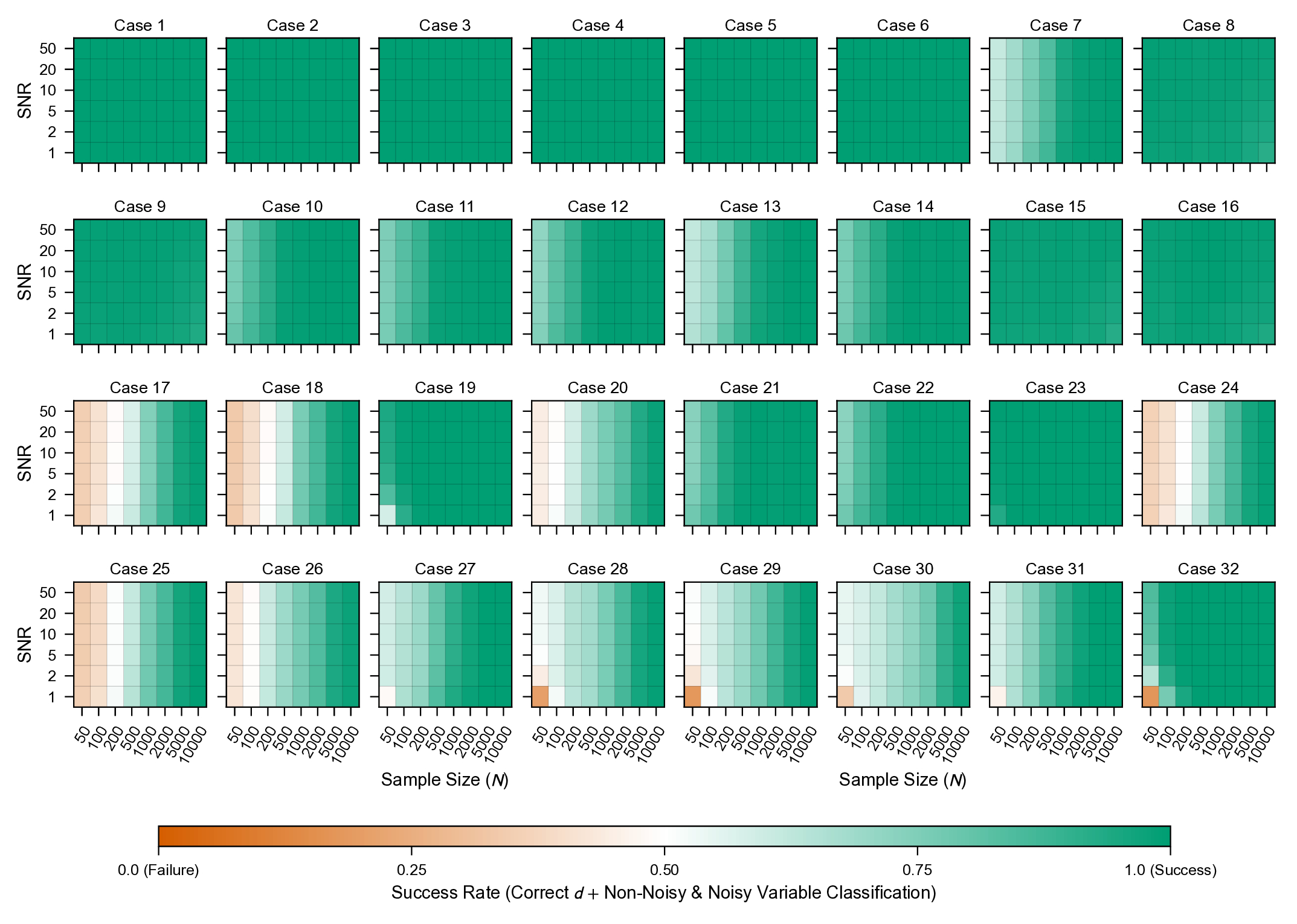}
  \caption{\textbf{Extended Data Fig. 4 $\mid$ Operating envelope of QZ-IPCA across all 32 noise configurations.} Each panel shows the success rate of QZ-IPCA over a grid of sample size $N \in \{50,$ $100,$ $200,$ $500,$ $1000,$ $2000,$ $5000,$ $10000\}$ (x-axis, log-scaled) and signal-to-noise ratios SNR $\in \{1,2,5,10,20,50\}$ (y-axis, log-scaled), for each of the 32 exhaustive noise configurations of the five-variable flow network (Figure~\ref{fig2:flow_network_cases}). Panels are ordered consistently with Figure~\ref{fig2:flow_network_cases}\textbf{b}. At each operating point $(N, \mathrm{SNR})$, $10^3$ independent Monte Carlo trials were performed. In each trial, $N$ true data samples of $F_1,F_2$ were generated by exciting the base flow values with sequences drawn from a Gaussian distribution with mean 0, and variances $\sigma^2_{F_1^*}=1$ and $\sigma^2_{F_2^*}=4$, respectively; these were subsequently used to generate the true values of the remaining flow variables. The true values were subsequently corrupted by additive Gaussian noise with variance fixed as $\sigma^2_{i}$ $=$ Var(true signal $F_i^*$)/SNR for each noisy variable $F_i$, computed once from the true data matrix and held fixed across all trials at that operating point. A trial is counted as a success only if all three conditions are simultaneously satisfied: (i) the number of deterministic constraints $d_{\text{non-noisy}}$ is correctly identified by PCA, (ii) the number of stochastic constraints $d_{\text{noisy}}$ is correctly identified via QZ-IPCA, and (iii) the noisy and non-noisy variable partition is correct. For variables in $n^{(2)}_{\text{non-noisy}}$, i.e., those not identified deterministically by PCA but estimated iteratively by QZ-IPCA, a variable is classified as non-noisy if its estimated variance satisfies $\hat{\sigma}^2<\left(0.1\times {\min}_{i:\sigma^2_{i}>0}\ \sigma^2_{i}\right)$, where the minimum is taken over all truly noisy variables in that case. The success rate at each operating point carries a $95\%$ confidence half-width of $\pm1.96 \times \sqrt{0.25/10^3} \simeq \pm3.1\%$. Color encodes success rate from $0.0$ (complete failure, vermillion) to $1.0$ (perfect success, bluish-green); the colormap is identical across all panels. Cases 1--16 achieve perfect or near-perfect success across the entire operating grid, with marginal degradation confined to $N\leq200$ in the most structurally complex cases among this group. Cases 17--32 reveal a well-defined phase boundary: success rate exceeds $0.5$ for all cases with $N>200$ regardless of SNR, and approaches $1.0$ for $N\geq2000$ and SNR $\geq5$, including Case 32 in which all five variables are noisy.}
\end{figure*}

\clearpage
\begin{figure*}[ht]
  \centering
  \includegraphics[width=\textwidth]{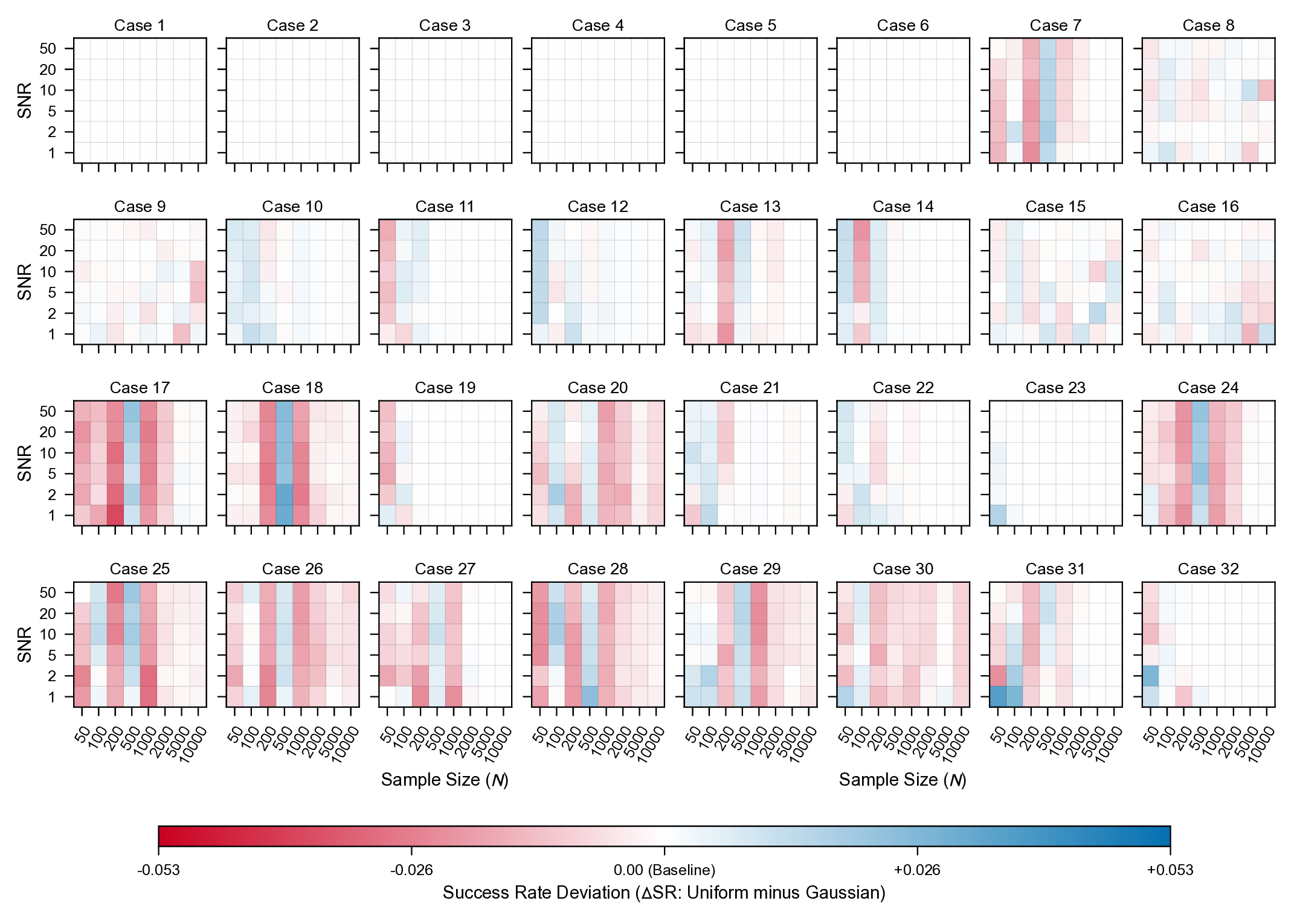}
  \caption{\textbf{Extended Data Fig. 5 $\mid$ Robustness of QZ-IPCA to sub-Gaussian excitation: success rate deviation from the Gaussian baseline.} Each panel shows the point-wise difference in the success rate, $\Delta \text{SR} = \text{SR}_{\text{Uniform}} - \text{SR}_{\text{Gaussian}}$, across the same ($N$,SNR) grid used in Extended Data Fig. 4, for each of the 32 noise configurations. The Gaussian baseline $\text{SR}_{\text{Gaussian}}$ is taken directly from Extended Data Fig. 4. In the non-Gaussian condition, the free variables $F_1$ and $F_2$ were excited using sequences drawn from a zero-mean uniform distribution on $[-\sqrt{3V_i}, +\sqrt{3V_i}]$, where $V_i=\sigma^2_{F_i^*}$ is the excitation variance of channel $i=\{1,2\}$, held identical to the Gaussian case. All other aspects of data generation, noise corruption, success criterion, and Monte Carlo procedure were unchanged ($10^3$ trials per operating point, $\pm 3.1\%$ CI half-width). Color encodes $\Delta \text{SR}$ on a diverging scale symmetric about zero, with red indicating underperformance ($\Delta \text{SR} < 0$) and blue indicating improvement ($\Delta \text{SR}>0$) relative to Gaussian; the colormap range and axis conventions are identical to Extended Data Fig. 6, permitting direct visual comparison between sub- and super-Gaussian excitation regimes. Cells with $|\Delta \text{SR}| \leq 0.031$ are indistinguishable from zero within Monte Carlo sampling noise and should be interpreted as consistent with the Gaussian baseline. Cases 1--16, which contain at least one deterministic constraint identified directly by PCA, show negligible deviation across the entire operating grid, confirming that the distributional assumption on excitation is inconsequential when deterministic structure governs the identification. In the fully stochastic cases (17--32 except cases 17 and 23 in which $d_{\text{non-noisy}}=1$),  deviations are confined to the finite-sample transition zone ($N=100-1000$) and show no consistent directional bias. Alternating blue and pink patches at intermediate $N$ reflect finite-sample covariance fluctuations rather than a systematic distributional disadvantage. The maximum observed deviation across all 32 cases and all 48 operating points is bounded within $|\Delta \text{SR}| \leq 0.053$ (5.3 percentage points), which, given the $\pm 3.1\%$ Monte Carlo noise floor, places the true worst-case signal at most 2.2 percentage points above the noise floor. All deviations converge to within the noise floor by $N=2000$ across all cases. These results confirm that QZ-IPCA achieves asymptotically equivalent identification performance under sub-Gaussian excitation as under Gaussian excitation, with only marginal and transient finite-sample deviations that are statistically indistinguishable from Monte Carlo noise at all but the most demanding operating points.}
\end{figure*}

\clearpage
\begin{figure*}[ht]
  \centering
  \includegraphics[width=\textwidth]{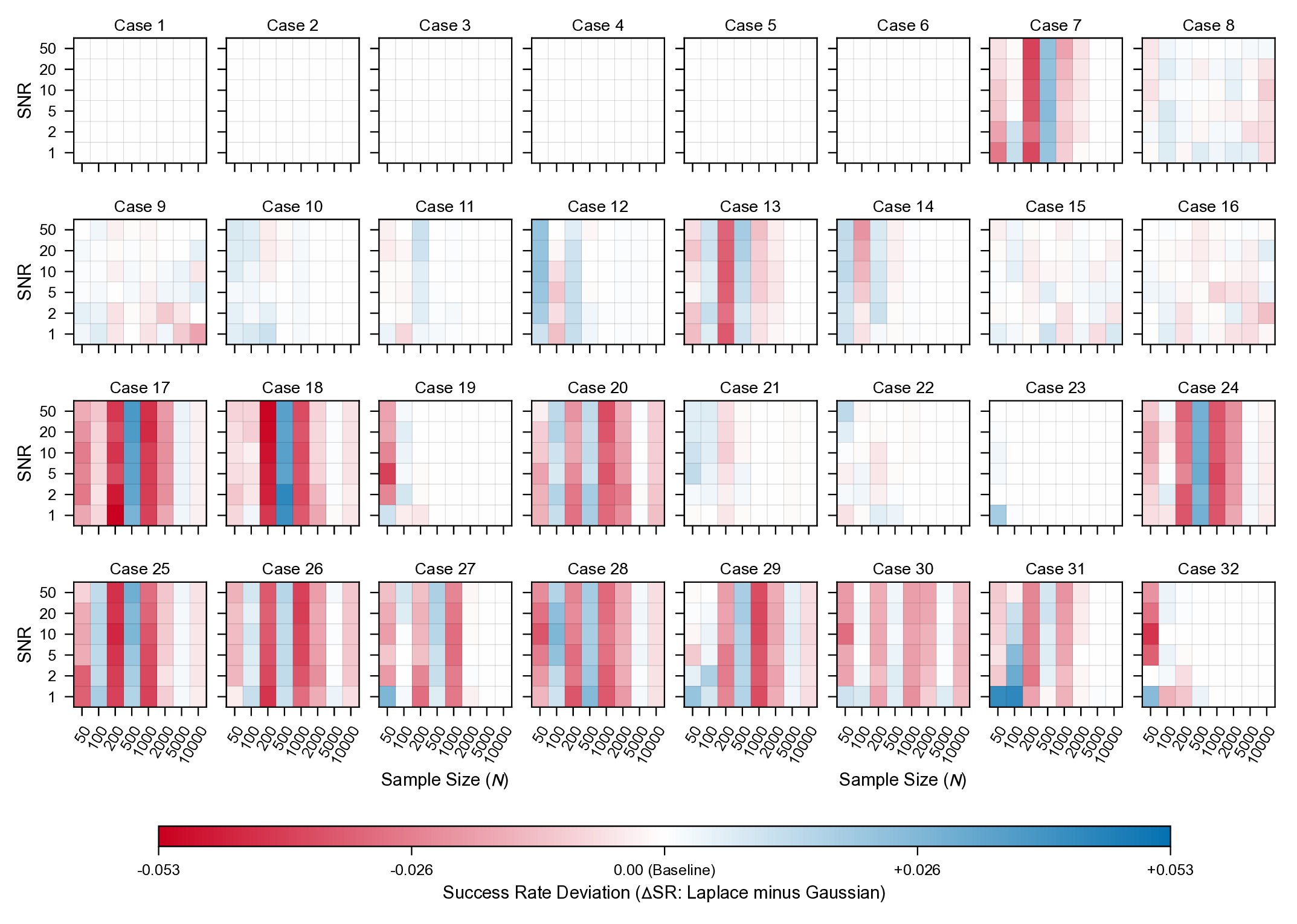}
  \caption{\textbf{Extended Data Fig. 6 $\mid$ Robustness of QZ-IPCA to super-Gaussian excitation: success rate deviation from the Gaussian baseline.} Each panel shows the point-wise difference in the success rate, $\Delta \text{SR} = \text{SR}_{\text{Laplace}} - \text{SR}_{\text{Gaussian}}$, across the same ($N$,SNR) grid used in Extended Data Fig. 4, for each of the 32 noise configurations. In the non-Gaussian condition, the free variables $F_1$ and $F_2$ were excited using sequences drawn from a zero-mean Laplace distribution with scale parameter $b_i = \sqrt{V_i/2}$, where $V_i=\sigma^2_{F_i^*}$ is the excitation variance of channel $i=\{1,2\}$, held identical to the Gaussian case, this parameterization preserves the excitation variance while introducing heavier tails (excess kurtosis of 3) relative to the Gaussian baseline. All other aspects of data generation, noise corruption, success criterion, and Monte Carlo procedure were unchanged ($10^3$ trials per operating point, $\pm 3.1\%$ CI half-width). The diverging colormap, symmetric axis range, panel ordering, and axis conventions are identical to Extended Data Fig. 3, permitting direct visual comparison between super- and sub-Gaussian excitation regimes. Cases 1--16 show negligible deviation across the entire operating grid, identical in character to the Uniform case (Extended Data Fig. 3), confirming that excitation distribution has no measurable effect (except cases 7 and 13 for $N=200$) on identification performance when deterministic constraints are present. In the fully stochastic cases (17--32), the Laplace condition produces a pattern of deviations qualitatively similar to the Uniform case but with slightly more saturated pink columns at intermediate $N$ in cases 17--18 and 24--30, consistent with the heavier tails of the Laplace distribution marginally slowing finite-sample convergence of the sample covariance matrix. Crucially, the maximum observed deviation across all 32 cases and all 48 operating points remains bounded within $|\Delta \text{SR}| \leq 0.053$ (5.3 percentage points), identical in magnitude to the Uniform case, placing both sub- and super-Gaussian excitation within the same tight robustness envelope. All deviations converge to within the $\pm 3.1\%$ Monte Carlo noise floor by $N=2000$ across as cases, and no residual bias is observed at large $N$. The near-identical worst-case bounds across Extended Data Figs. 5 and 6 establish that the robustness of QZ-IPCA is insensitive to whether the excitation distribution has lighter or heavier tails than Gaussian: identification performance is asymptotically equivalent in all three distributional settings, with finite-sample deviations bounded by 5.3 percentage points and confined to the sample-size transition zone.}
\end{figure*}

%
\clearpage
\begin{figure*}[ht]
  \raggedright
  {\small\bfseries Extended Data Table 1 $\mid$ Noisy variable identification and variance estimation for cases 1--16}\par
  \vspace{1mm} 
  \centering
  \includegraphics[width=\textwidth]{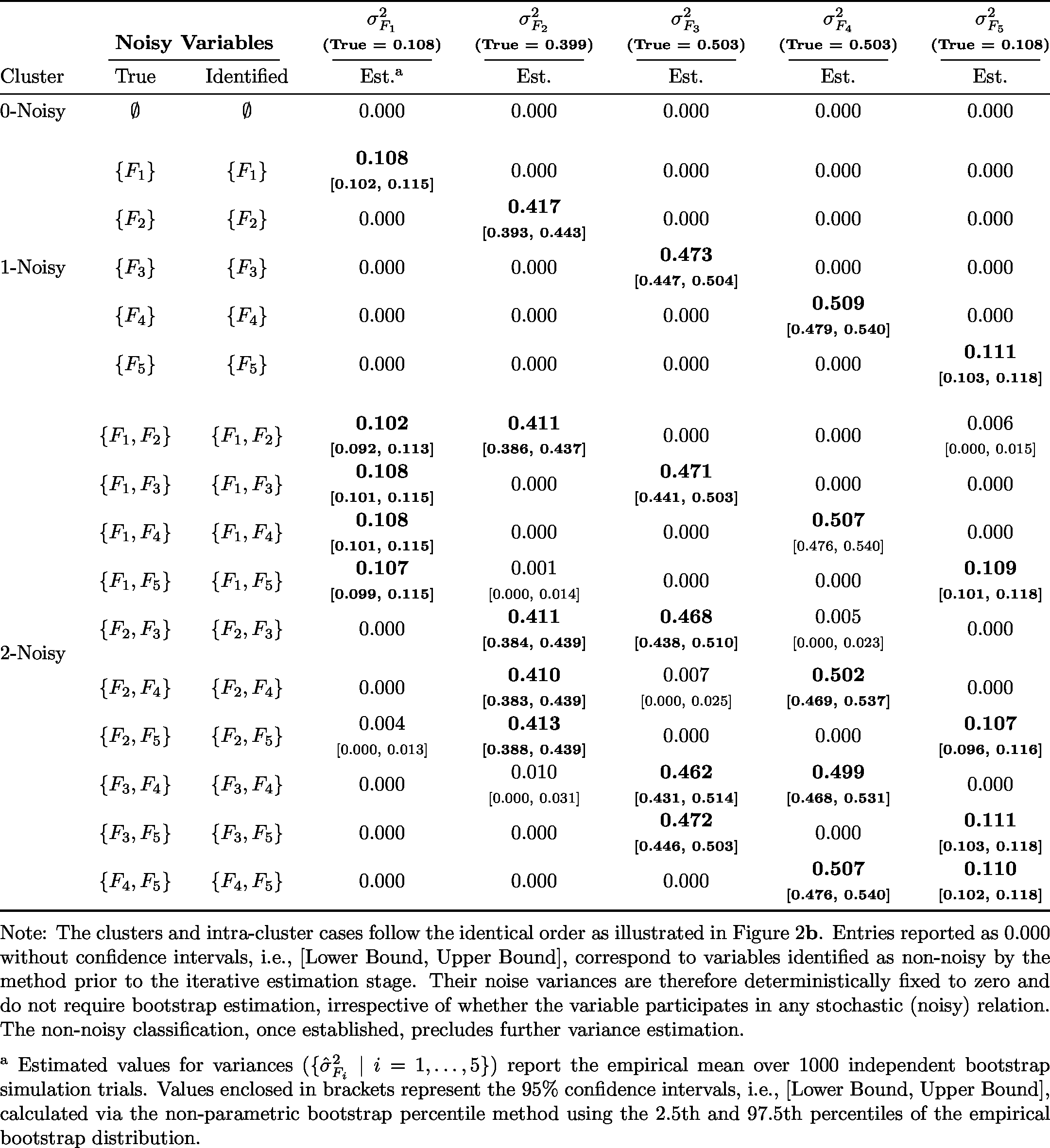}
\end{figure*}

\clearpage
\begin{figure*}[ht]
  \raggedright
  {\small\bfseries Extended Data Table 2 $\mid$ Noisy variable identification and variance estimation for cases 17--32}\par
  \vspace{1mm} 
  \centering
  \includegraphics[width=\textwidth]{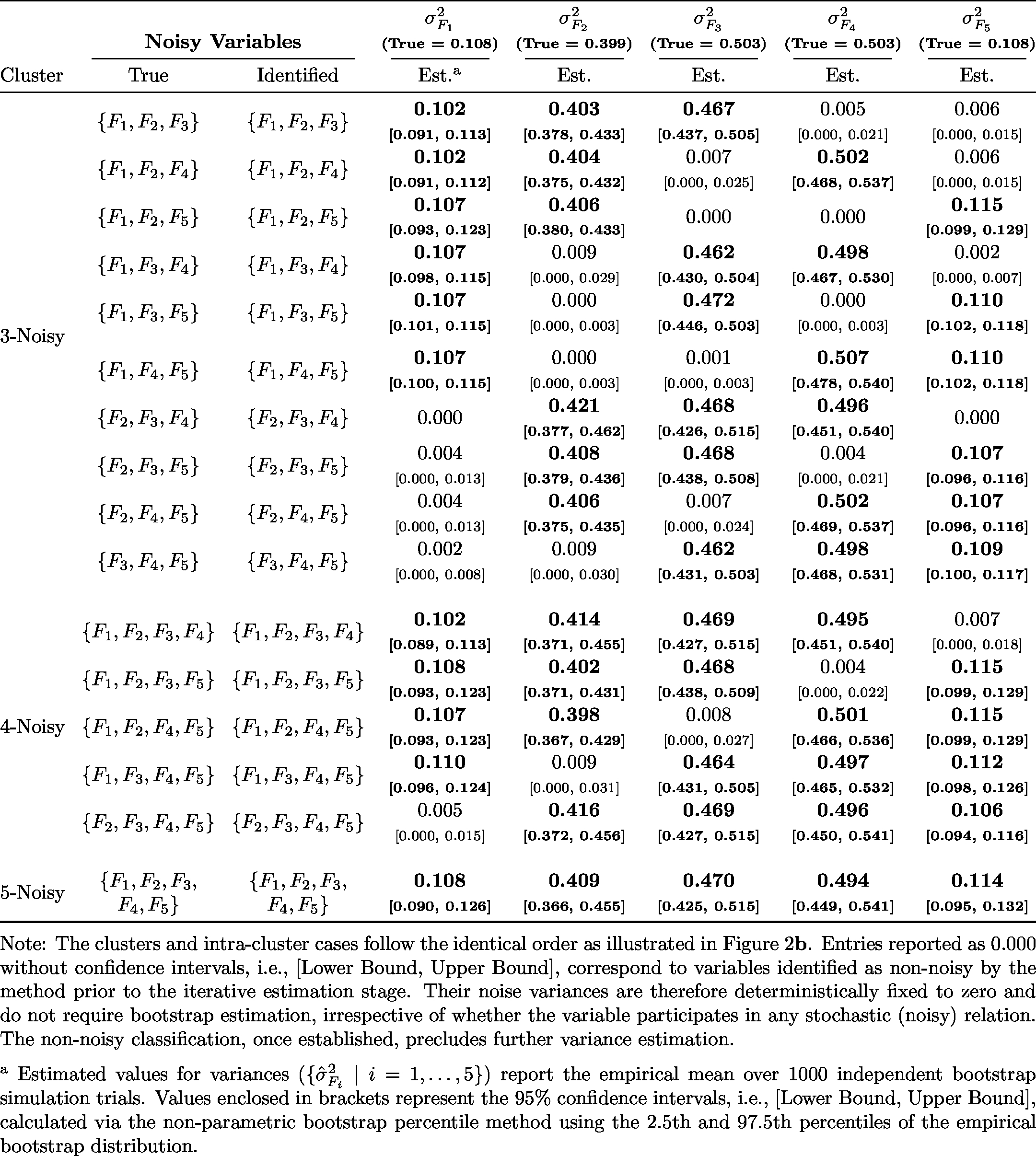}
\end{figure*}

\clearpage
\begin{figure*}[ht]
  \raggedright
  {\small\bfseries Extended Data Table 3 $\mid$ Model structure identification and coefficient recovery}\par
  \vspace{1mm} 
  \centering
  \includegraphics[width=\textwidth]{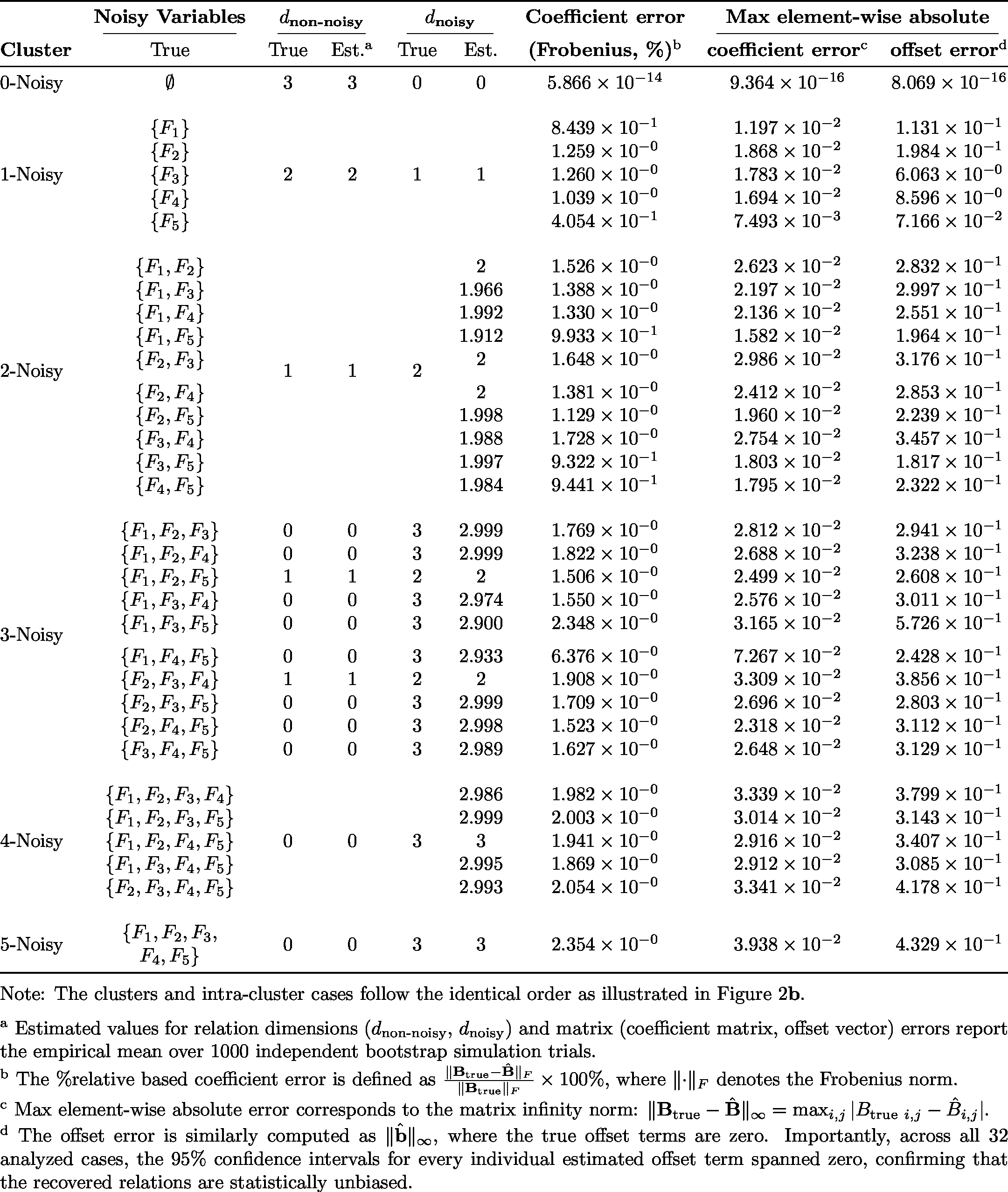}
\end{figure*}

\end{document}